\documentclass[11pt]{article}

\usepackage{graphicx}			
\usepackage[font=small]{subcaption}			
\usepackage{flafter}			
\usepackage{amsmath}						
\usepackage{amssymb}
\usepackage{array}							
\newcolumntype{?}[1]{!{\vrule width #1}}	
\usepackage{booktabs} 
\usepackage{amsfonts} 
\usepackage[noabbrev]{cleveref} 
\usepackage{bbm} 
\usepackage[a4paper, total={6.5in, 9in}]{geometry} 
\usepackage{caption}
\usepackage{cleveref}
\usepackage{float}  

\DeclareMathOperator\erf{erf}
\DeclareMathOperator{\sech}{sech}

\begin{document}
	\begin{titlepage}
		
		\vspace*{-1cm}
				
		\vskip 3cm
				
		\vspace{.2in}
		\begin{center}
			{\large\bf Aspects of the 
				modified regularized long-wave equation}
		\end{center}
		
		\vspace{.5cm}
		
		\begin{center}
			F. ter Braak$^{\dagger}$ and W. J. Zakrzewski$^{\dagger \dagger}$
					
			\vspace{.5 in}
			\small
			
			\par \vskip .2in \noindent
			Department of Mathematical Sciences,\\
			University of Durham, Durham DH1 3LE, U.K.\\ 
			${}$ \newline 			
			$^{(\dagger)}$ floris.ter-braak@durham.ac.uk \\
			$^{(\dagger \dagger)}$ w.j.zakrzewski@durham.ac.uk
			
			\normalsize
		\end{center}
		
		\vspace{.5in}	
	
		\begin{abstract}
			We study various properties of the soliton solutions of the modified regularized long-wave equation. This model possesses exact one- and two-soliton solutions but no other solutions are known. We show that numerical three-soliton configurations, for which the initial conditions were taken in the form of a linear superposition of three single-soliton solutions, evolve in time as three-soliton solutions of the model and in their scatterings each individual soliton experiences a total phase-shift that is the sum of pairwise phase-shifts. We also investigate the soliton resolution conjecture for this equation, and find that individual soliton-like lumps initially evolve very much like lumps for integrable models but eventually (at least) some blow-up, suggesting basic instability of the model.
		\end{abstract}
	\end{titlepage}
	
	\section{Introduction} \label{Introduction}
	
	Integrable partial differential equations in~$(1+1)$ dimensions, such as the KdV and the non-linear Schr\"{o}dinger equation, possess infinitely many conservation laws which implies that such systems possess soliton solutions; these solitons are localised waves which preserve their shape and velocity before and after the scattering, but they do experience a phase-shift as a result of the scattering~\cite{Hietarinta0}. However, integrable models are quite rare, and some physical events can be described by models which possess soliton-like structures (such as general vortices, skyrmions and baby skyrmions) but are not integrable. Furthermore, the scattering properties of their soliton-like solutions are often not that different, in the sense that the amount of emitted radiation is not very large. Therefore, these models can loosely be described as `almost-integrable'. This has lead to various attempts to define the concept of \textit{quasi-integrability} (see, {\it e.g.}, paper~\cite{Ferreira} and references therein). Attempts have also been made to relate this concept to the extra (very special) symmetries satisfied by the two-soliton solutions~\cite{Ferreira1}.  
		
	In fact, the whole concept of integrability can be defined in various ways~\cite{LMS}. 
 One of such definitions is \textit{Hirota integrability}, which is based on Hirota's method for
 obtaining multi-soliton solutions of non-linear models (and we will, when it is important
to stress this fact, refer to them as Hirota solutions). In addition, it is often claimed that if this 
method leads to the construction of the exact one-, two- and three-soliton Hirota solutions, the model is 
considered to be Hirota integrable~\cite{Hietarinta}. On the other hand, partial differential equations 
which only possess one- and two-soliton solutions, also known as \textit{partial integrable} models, 
are not Hirota integrable~\cite{Hietarinta0}. (Note that this does not necessarily mean that solutions 
describing three or more solitons do not exist for these models; it only shows that they cannot 
be obtained using the Hirota method.) One of such models is the modified regularized long-wave (mRLW)
 equation,\footnote{We thank J. Hietarinta for drawing our attention to this model.} which we discuss in detail in this paper. 
		
	It has also been observed that if a set of equations is Hirota integrable, they satisfy the more 
conventional definitions of integrability~\cite{Hietarinta1}. Thus it is interesting to try to better understand why, whereas models with only one- and two-soliton Hirota solutions do not appear to satisfy the conventional definitions of integrability, systems which also possesses three-soliton Hirota solutions do satisfy the conventional definitions of integrability. This raises the question: what is so special about three-soliton Hirota solutions?
		
	In this paper we look in detail at various (numerical) properties of the mRLW equation. 
In the next section we recall this equation and the exact form of its one- and two-soliton solutions. Then we
 compare the analytical two-soliton scattering with our numerical approximation in order to test the accuracy 
and stability of our finite difference scheme. We also check whether the numerically evolved linear superposition of two single-soliton solutions is a 
good approximation to the corresponding analytical two-soliton solution,
 because we want to use a linear superposition of three single-soliton solutions as the initial conditions 
for our numerical three-soliton simulations. We find 
that this approach gives us essentially the same results as the ones obtained by using the exact
 two-soliton solutions. In section~\ref{Numerical_three_soliton_solution}
 we investigate the time evolution of three-soliton systems obtained in such a way and 
look at the phase-shifts experienced by these solitons during their interactions.  Finally,
 in section~\ref{soliton_resolution_conjecture} we look at some simulations using lumps ({\it i.e.}, 
functions which do not solve the mRLW equation) in order to test the soliton resolution conjecture. 
The last section of the paper presents our conclusions and plans for future work. 
	
	A large part of our results is based on numerical simulations of the mRLW equation. Since we used a numerical procedure which combined explicit and implicit finite difference methods, the work involved the discretisation in both time and space, and so had large memory requirements. Most of the simulations involved grid-spacing of~$h=0.1$ and time-steps of~$\tau = 0.001$. To assess the reliability of our procedures we have altered these values and we are confident that the results presented in the paper are correct. Since our procedure is second order in time, we require an analytic expressions of the fields at the first $2$ time levels. We took them from the expressions of $q(x,t)$ given in the paper and refer to as `initial conditions'. The spatial
	boundary conditions were fixed. More details on the numerical procedure are presented in appendix~\ref{Numerical_approx},
	and the particular values of various parameters used in the reported simulations are gathered in a table in appendix~\ref{summary_of_variables}.
	
	\section{The modified regularized long-wave equation} \label{Modified_regularized_long-wave_equation_equation}
		
	The mRLW equation, introduced by J. D. Gibbon, J. C. Eilbeck and R. K. Dodd~\cite{Gibbon}, is defined by
	\begin{equation}
		q_{xxtt} + 2q_{xx} q_{tt} + 4q_{xt}^2 - q_{xt} - q_{tt} = 0 \,, \label{MRL}
	\end{equation}
	where~$q = q(x, t)$ is a real-valued function, and the subscripts~$x$ and~$t$ denote partial differentiation with respect to these variables. It is known that this equation possesses analytical one- and two-soliton Hirota solutions~\cite{Gibbon}, where the one-soliton solutions are given by
	\begin{equation}
		q = \ln \left(1 + e^{\eta_1} \right) \label{1.1}
	\end{equation}
	and the two-soliton solutions take the form 
	\begin{equation}
		q = \ln \left( 1 + e^{\eta_1} + e^{\eta_2} + A_{12} e^{\eta_1 + \eta_2} \right) \,, \label{1.2}
	\end{equation}
	where
	\begin{equation}
		\eta_i = k_i x - \omega_i t + \delta_i \,,\quad i = 1,2 \,,
	\end{equation}
	and
	\begin{equation}
		A_{12} = - \frac{(\omega_1 - \omega_2)^2(k_1 - k_2)^2 + (\omega_1 - \omega_2)(k_1 - k_2) - (\omega_1 - \omega_2)^2}{(\omega_1 + \omega_2)^2(k_1 + k_2)^2 + (\omega_1 + \omega_2)(k_1 + k_2) - (\omega_1 + \omega_2)^2} \,. \label{1.2.1}
	\end{equation}
	In these expressions, the parameters~$k_i$ and~$\omega_i$ are constrained by the following dispersion relation
	\begin{equation}
		\omega_i = \frac{k_i}{1-k_i^2} \,, \quad i = 1,2 \,. \label{1.3}
	\end{equation}
	The actual soliton fields of Gibbon \textit{et al.\ }are defined by $u \equiv - q_{xt}$.

\begin{figure}[b!]
		\centering
		\hspace*{-0.1cm}
		\begin{subfigure}{.34\textwidth}
			\centering
			\includegraphics[scale=0.28]{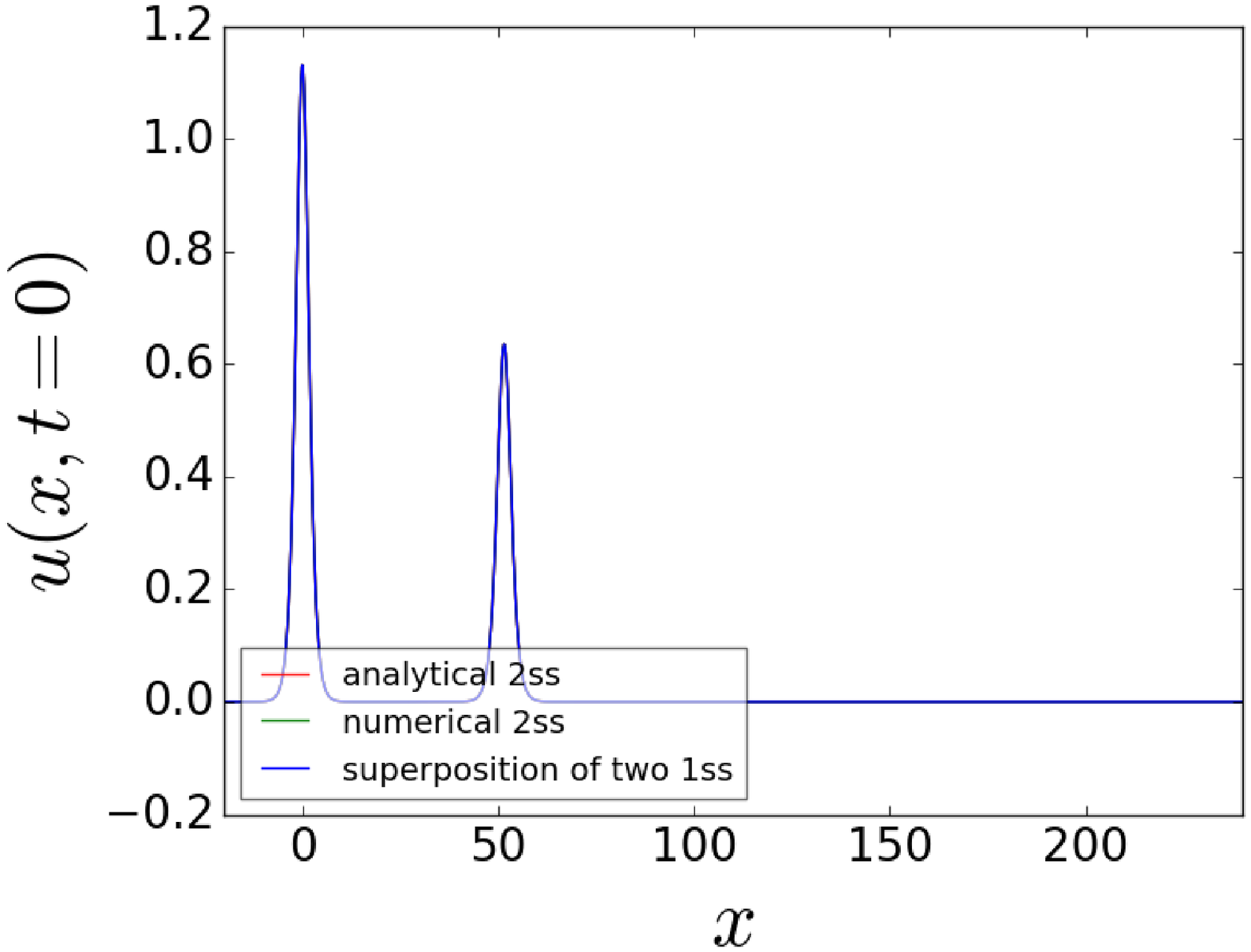}
			\caption{At~$t=0$}
			\label{plot0_2}
		\end{subfigure}%
		\begin{subfigure}{.34\textwidth}
			\centering
			\includegraphics[scale=0.28]{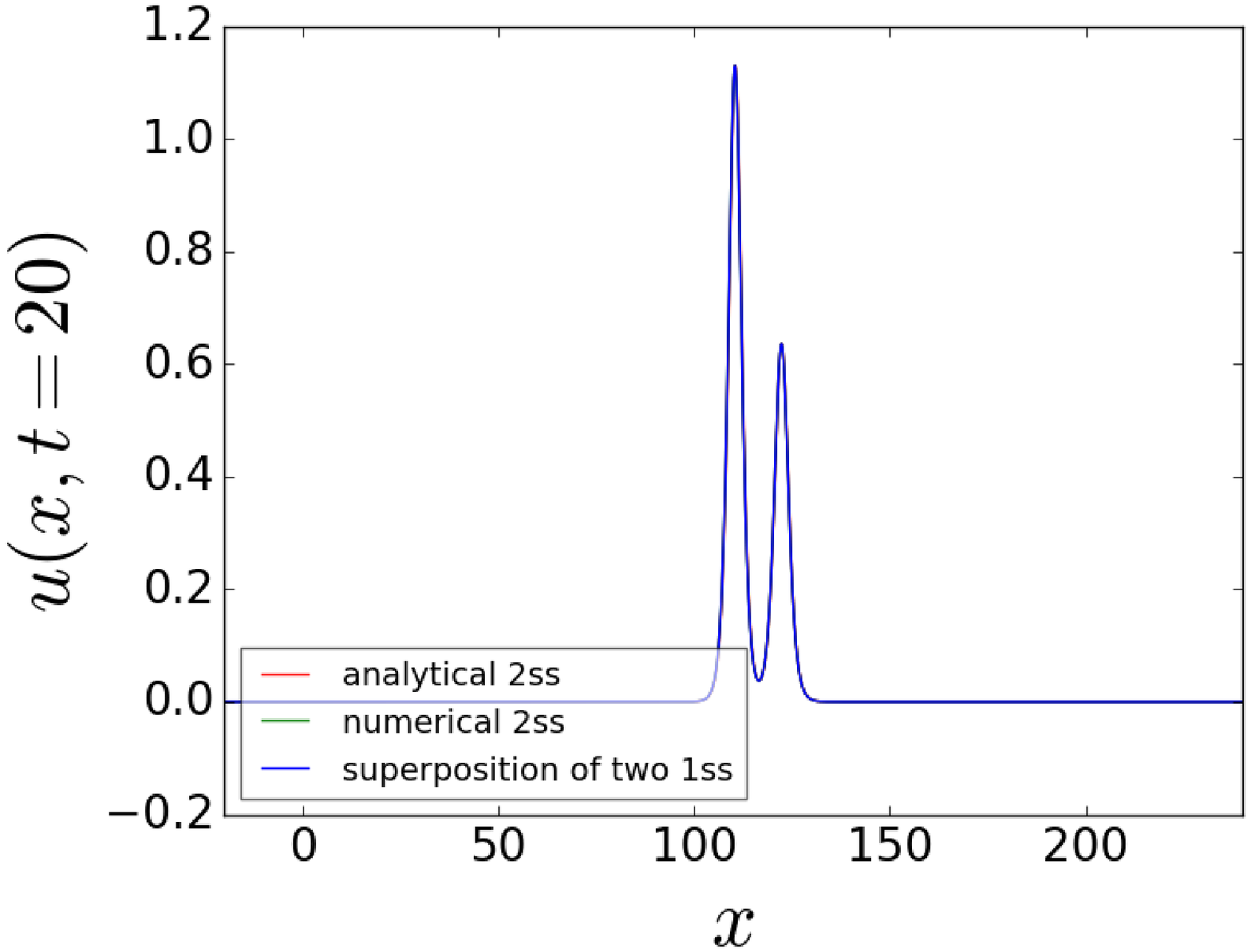}
			\caption{At~$t=20$}
			\label{plot0_3}
		\end{subfigure}%
		\begin{subfigure}{.34\textwidth}
			\centering
			\includegraphics[scale=0.28]{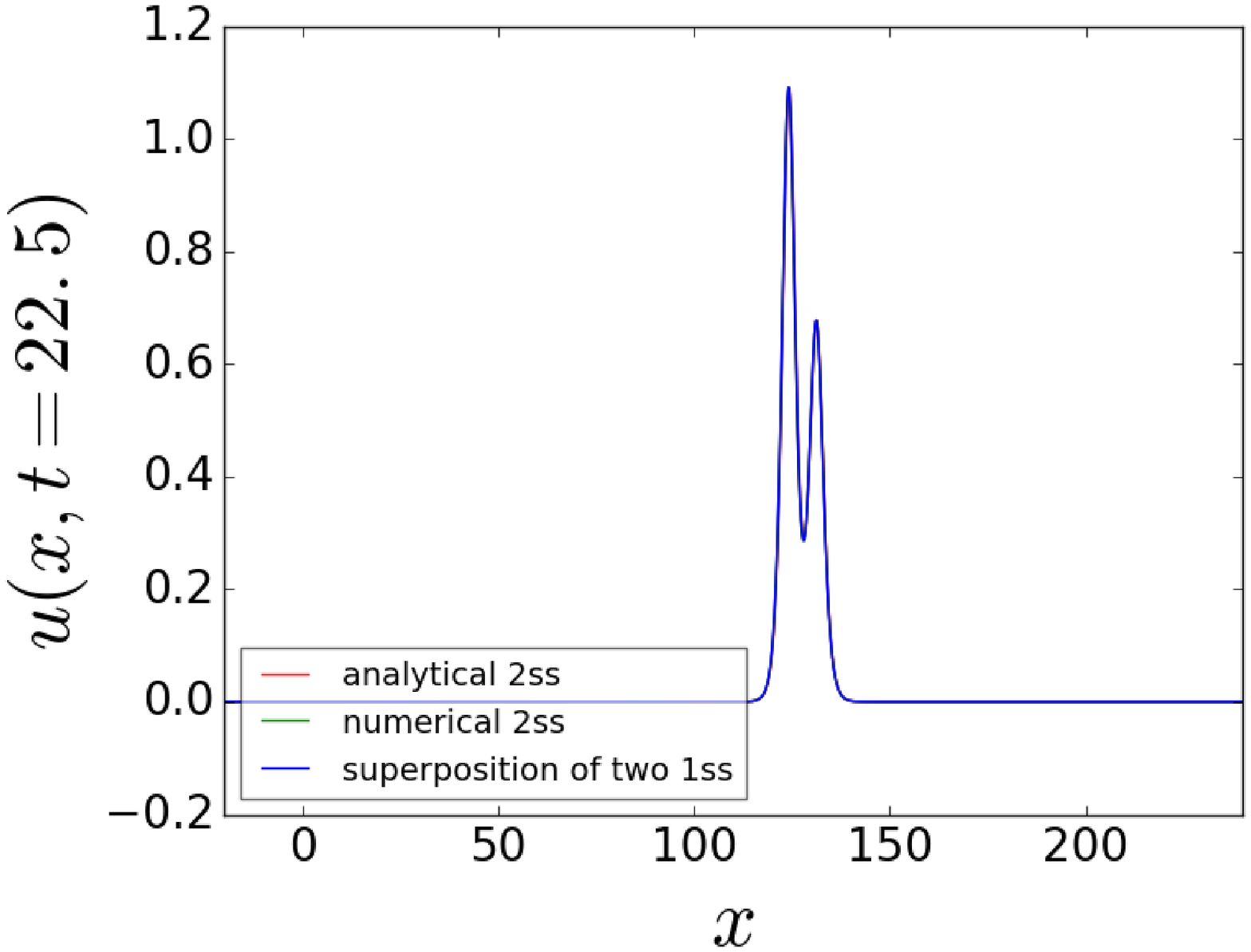}
			\caption{At~$t=22.5$}
			\label{plot0_4}
		\end{subfigure}
		
		\centering
		\hspace*{-0.1cm}
		\begin{subfigure}{.34\textwidth}
			\centering
			\includegraphics[scale=0.28]{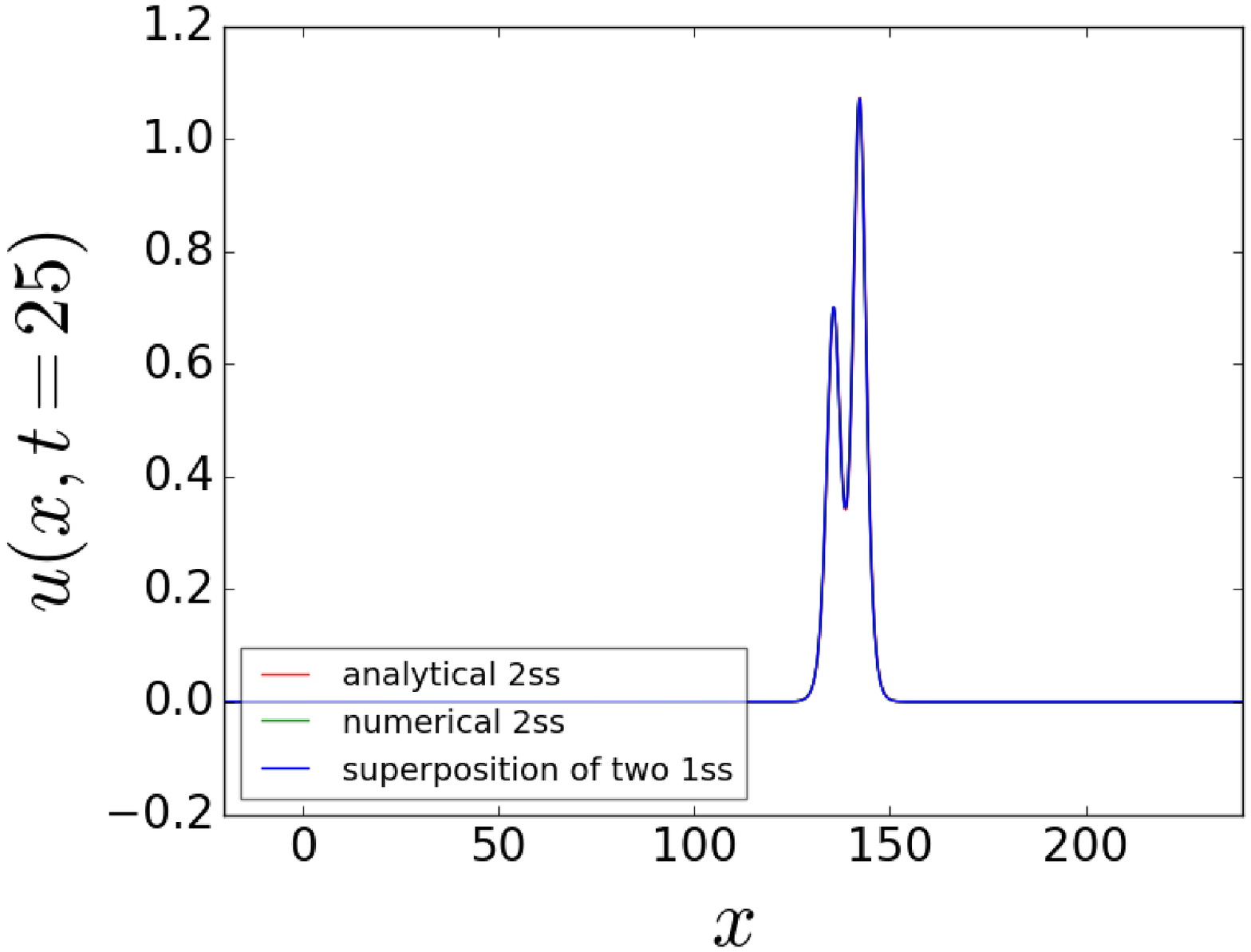}
			\caption{At~$t=25$}
			\label{plot_5}
		\end{subfigure}%
		\begin{subfigure}{.34\textwidth}
			\centering
			\includegraphics[scale=0.28]{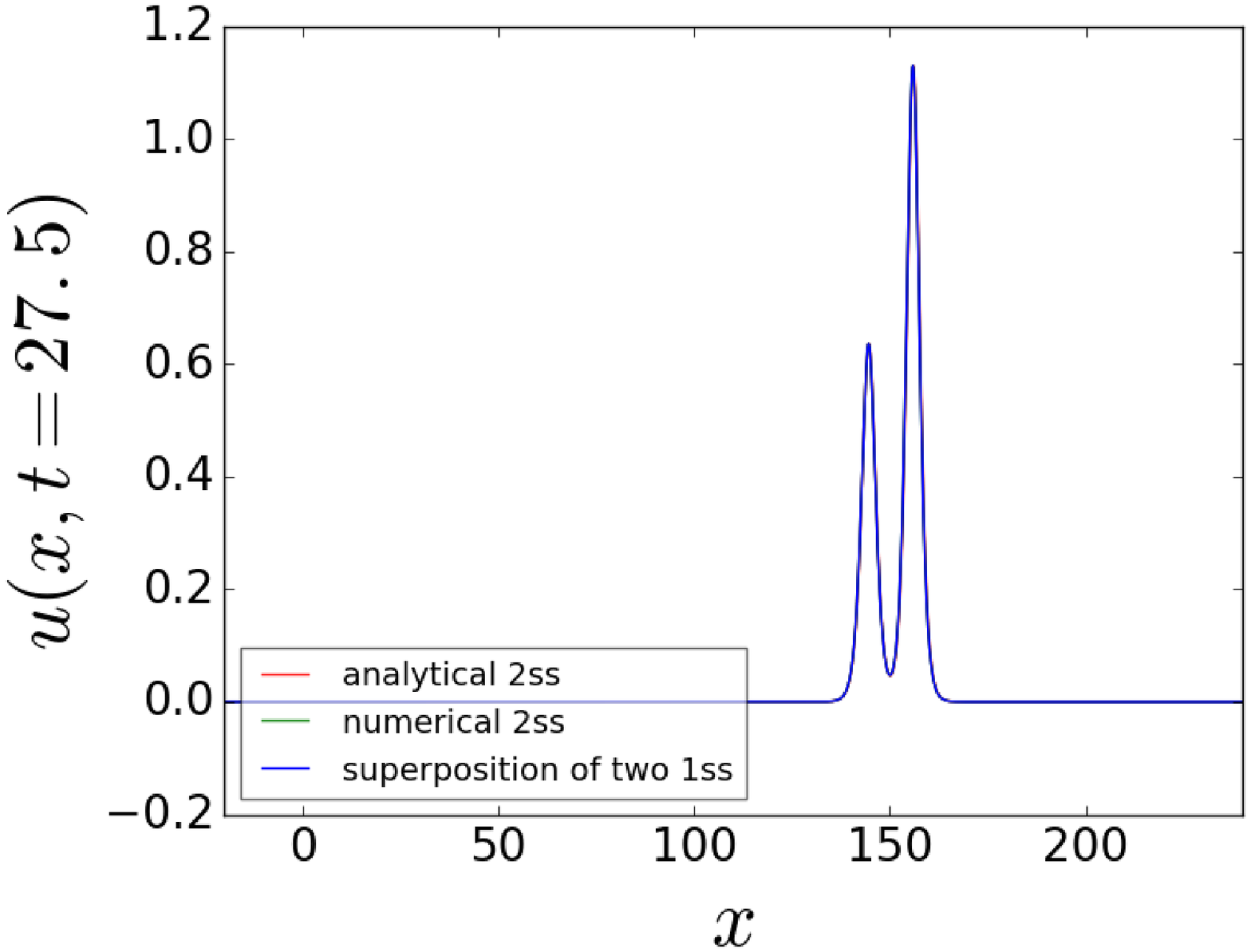}
			\caption{At~$t=27.5$}
			\label{plot0_6}
		\end{subfigure}%
		\begin{subfigure}{.34\textwidth}
			\centering
			\includegraphics[scale=0.28]{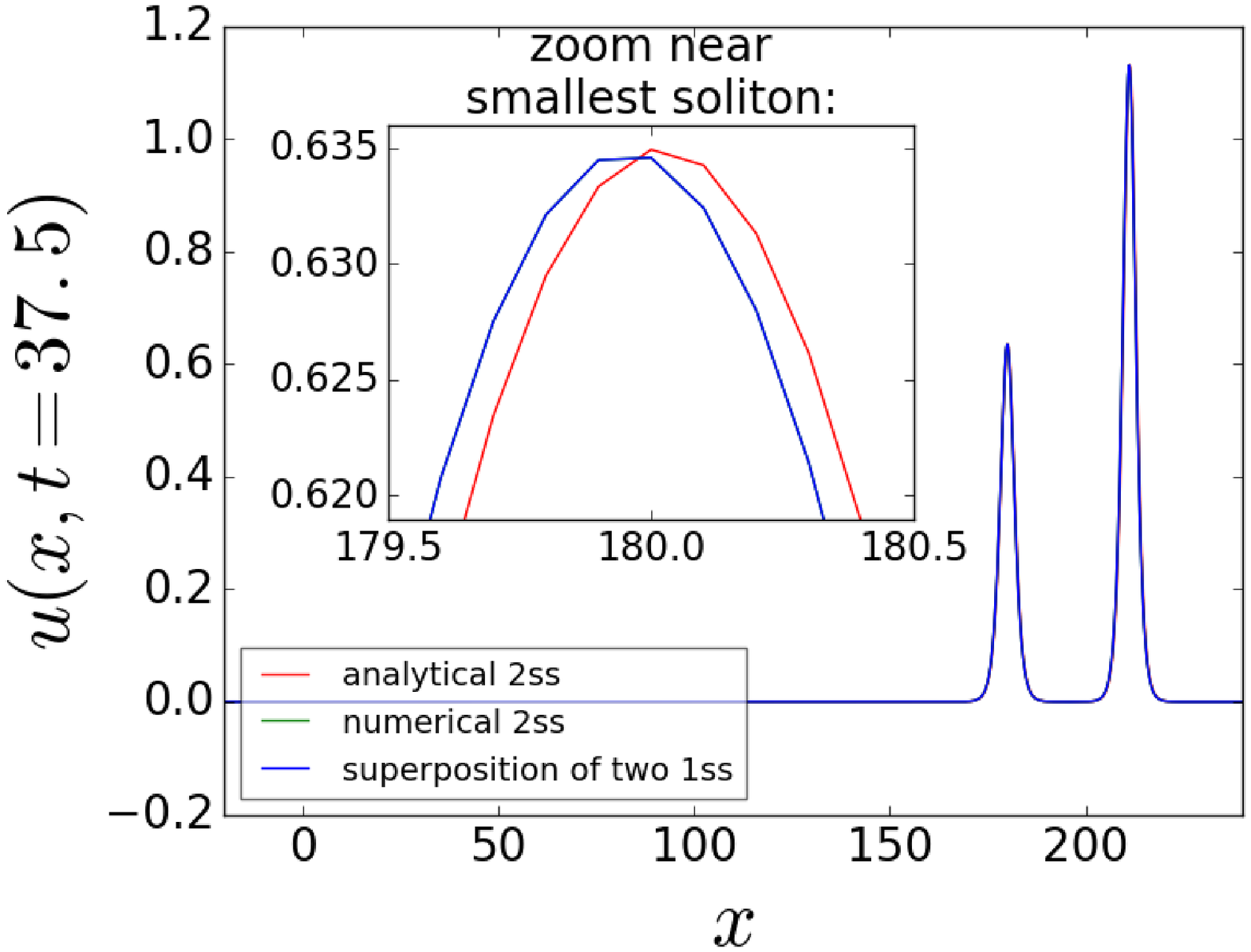}
			\caption{At~$t=37.5$}
			\label{plot0_7}
		\end{subfigure}
		\caption{This figure shows two solitons interacting with each other at different points in time for three simulations. The red line is obtained using the analytic expression given by equation~(\ref{1.2}), the green line shows the numerical time evolution of equation~(\ref{1.2}) as initial conditions, and the blue line shows the numerical time evolution of equation~(\ref{1.4}) as initial conditions. (Note that the three simulations are so close that their plots are barely distinguishable with the naked eye.)} 
		\label{plot0_2to0_7}
	\end{figure}

	In what follows, for definiteness, we consider only~$1 > k_1 > k_2 > 0$. This implies that
 both the amplitude and velocity of each soliton will be positive. Furthermore, the soliton corresponding 
to~$\eta_1$ will have a larger amplitude and velocity than the soliton corresponding to~$\eta_2$.
 To illustrate this, in figure~\ref{plot0_2to0_7} the red lines present the plots of the spatial 
dependence of $u$ at various values of $t$ of an analytical two-soliton simulation. The other curves will be discussed below.

	The Hirota method does not generate three-soliton solutions for the mRLW equation,
 which implies that the model is not Hirota integrable. Furthermore, to our knowledge,
 nobody has found an analytic expression describing three or more solitons (using any method).
 Thus, we need to use numerical methods (see the appendix for more details) in order to study the 
time evolution of three-soliton configurations. To test this scheme empirically, we have used the initial
 field configurations ({\it i.e.}, the initial conditions) expressed by equation~(\ref{1.2}) and evolved
the configuration in time using our procedures. The green lines in figure~\ref{plot0_2to0_7} 
shows the plots for such a simulation which is produced using the same values for the angular frequencies
 $\omega_i$ and phase constants $\delta_i$ as for the aforementioned analytical simulation (see appendix~\ref{summary_of_variables} for more information).
 This allows us to compare the numerical solution with the analytical results. Since these two lines in figure~\ref{plot0_2to0_7} are so close to each other that one can hardly distinguish them, the numerical solution is a good approximation of the analytical values. 
	
	Furthermore, it will be useful for the simulations discussed in the next section to also consider the time evolution of the following initial conditions
	\begin{equation}
		q = \ln \left(1 + e^{\eta_1} \right) + \ln \left(1 + e^{\eta_2} \right) \,.  \label{1.4}
	\end{equation}
	The blue lines in figure~\ref{plot0_2to0_7} present the results of this evolution,
 where again the parameters governing $\eta_1$ and $\eta_2$ have exactly the same values as for 
the previously two discussed simulations. All together, figure~\ref{plot0_2to0_7} demonstrates that the
 numerical time evolution of a 
linear superposition of two exact one-soliton solutions produces results that are also almost 
indistinguishable from those of the analytical and numerical two-soliton simulations (provided that the two solitons are initially placed far apart from each other).
			
	The above described three different simulations are so close that with the naked eye they are 
indistinguishable on the scale used in figure~\ref{plot0_2to0_7}. Therefore, in order to illustrate how close
 the three lines are, in figure~\ref{plot0_7} we have added an insert of the region near the amplitude of the smallest
 soliton on a much smaller scale. This insert demonstrates very clearly that there is indeed a very small
 discrepancy between the analytical result and both the numerical simulations, but even on this 
scale we cannot distinguish between the numerical time evolution of equation~(\ref{1.2}) 
and equation~(\ref{1.4}) (and upon zooming in on the region of the larger soliton, we find 
that the discrepancies between the simulations are of similarily small magnitudes). This is expected since looking at the time evolution of the two solitons, we see that most of the interaction takes place when the solitons are close together. We see that in this case each soliton has a size of about~$L\sim 20$, and since they are initially placed further apart than~$L$ (see figure~\ref{plot0_2}), the errors of such an approximation are only in the interaction of their `tails'. 

	When the two solitons scatter, the only result of their interaction after the collision is the phase-shift they experience. In order to determine the analytical expression of this phase-shift, let us introduce two new variables $y \equiv x - \frac{\omega_1}{k_1}t$ and $z \equiv x - \frac{\omega_2}{k_2}t$. Then, substituting~$y$ into~$\eta_1$ and~$\eta_2$ ({\it i.e.},~$\eta_1 = k_1 y + \delta_1$ and~$\eta_2 = k_2(y + \frac{\omega_1}{k_1} t) - \omega_2t + \delta_2$ 
	) the exact two-soliton solution can be asymptotically approximated as
	\begin{equation}
		\lim_{t \to - \infty} u(y,t) \approx \frac{k_1 \omega_1 e^{\eta_1}}{\left( 1 + e^{\eta_1} \right)^2} \quad \text{and} \quad \lim_{t \to \infty} u(y,t) \approx \frac{k_1 \omega_1 e^{\eta_1 + \ln A_{12}}}{\left( 1 + e^{\eta_1 + \ln A_{12}} \right)^2} \,.
	\end{equation}
	Similarly, introducing the variable~$z$ gives
	\begin{equation}
		\lim_{t \to - \infty} u(z,t) \approx \frac{k_2 \omega_2 e^{\eta_2 + \ln A_{12}}}{\left( 1 + e^{\eta_2 + \ln A_{12}} \right)^2}  \quad \text{and} \quad \lim_{t \to \infty} u(z,t) \approx \frac{k_2 \omega_2 e^{\eta_2}}{\left( 1 + e^{\eta_2} \right)^2} \,.
	\end{equation}
	Thus, after the collision the solitary wave corresponding to~$\eta_1$ is phase-shifted forward by~$\ln A_{12}$ and the wave corresponding to~$\eta_2$ is phase-shifted by~$\ln A_{12}$ in the opposite direction.
		
	In the following section we analyse the phase-shift that solitons experience during three-soliton
 scattering. Since no three-soliton solutions are known, we investigate them numerically, 
and so it will be useful to test the reliability of our numerical method for determining the phase-shifts
 of the analytical and numerical two-soliton simulations.\footnote{We determine the phase-shift
 by finding the three highest points of each individual soliton at some $t$, and assume they fit a
 polynomial of degree~$2$. Subsequently, we use each polynomial to estimate the position and height
 of its absolute maximum. We repeat this procedure for many values of time (see, for instance, 
figures~\ref{1amplitude_vs_time_and_1location_vs_time} and~\ref{3amplitude_vs_time_and_1location_vs_time}
 in the next section), and this has allowed us to determine the phase-shift experienced by the solitons.}
		We estimate this error by dividing the
 computationally observed phase-shift (from either the analytical or numerical simulation)
 by $\ln A_{12}$  ({\it i.e.}, the analytical expression of the phase shift-shift). 
With this definition, we have found that the error is always small but it is somewhat sensitive
to the details of our procedure. In fact we have found that in all our (analytical and numerical) simulations 
 the error had always been smaller than $5.0 \%$. Since we also observed errors close to $5.0 \%$ for 
the analytical simulations, we can conclude that these errors are mainly a result 
of the algorithm of determining the phase-shift rather than resulting from the finite
 difference scheme approximating the mRLW equation.

	Finally, let us briefly discuss the conserved charges for the mRLW equation. As far as we are aware, its only known conservation laws are given by
	\begin{equation}
		\partial_t \int\limits^\infty_{-\infty } \mathrm{d} x \;  (-q_{xt}) = \partial_t \int\limits^\infty_{-\infty } \mathrm{d} x \;  u = 0 
	\end{equation}
	and
	\begin{equation}
		\partial_t \int\limits^\infty_{-\infty } \mathrm{d} x \;  (-q_{xx}) = 0 \,,
	\end{equation}
	which can be easily verified by taking the derivative of equation~(\ref{MRL}) with respect to $x$. Let us add that one can approximate the values of these conserved charges for the analytical one- and two-soliton solutions as follows
	\begin{equation}
		Q_1 \equiv \int\limits^\infty_{-\infty } \mathrm{d} x \;  (-q_{xt}) = [-q_t]^\infty_{x=-\infty} \approx \sum\limits_{i=1}^N \omega_i  \,,\quad N = 1,2 \,,  \label{1.5}
	\end{equation}
	and
	\begin{equation}
		Q_2 \equiv \int\limits^\infty_{-\infty } \mathrm{d} x \;  (-q_{xx}) = [-q_x]^\infty_{x=-\infty} \approx -\sum\limits_{i=1}^N k_i  \,,\quad N = 1,2 \,. \label{1.6}
	\end{equation}
	In the next section, we check whether these quantities are also conserved 
for the numerically evolved three-soliton configurations.

	\section{Numerical three-soliton solutions} \label{Numerical_three_soliton_solution}
	
	So far we discussed mainly two-soliton configurations. In this section we
 look at systems involving three solitons. As we stated before, the analytical three-soliton
 solutions (or solutions involving even more solitons) of the mRLW equation are not known, and 
they cannot be found by the Hirota method. This has been stated in literature and we have verified 
this claim for three and four solitons. So, we do not really know whether such solutions exist or not; all 
we know is that if they exist, their forms cannot be found by the Hirota method. 
	
	As we have shown in the previous section, a linear superposition of two single-soliton solutions 
is almost indistinguishable from the analytical two-soliton solution when, initially, these two 
solitons are far enough apart. Armed with this observation, we have numerically 
simulated the time evolution of a three-soliton system by using the linear superposition
 of three well-separated one-soliton solutions as the initial conditions for our simulations. 
In other words, we use
	\begin{equation}
		q = \ln \left(1 + e^{\eta_1} \right) + \ln \left(1 + e^{\eta_2} \right) + \ln \left(1 + e^{\eta_3} \right) \label{sup3}
	\end{equation}
	to construct the initial conditions of a three-soliton system. 
  We have performed many such simulations using a range of variables describing the frequencies of 
individual solitons and, in the next subsection,  we discuss the results of two of such simulations for illustrative purposes. 
		
	\subsection{Three-soliton interactions}
	
	
	
\begin{figure}[t!]
		\centering
		\hspace*{-0.1cm}
		\begin{subfigure}{.34\textwidth}
			\centering
			\includegraphics[scale=0.28]{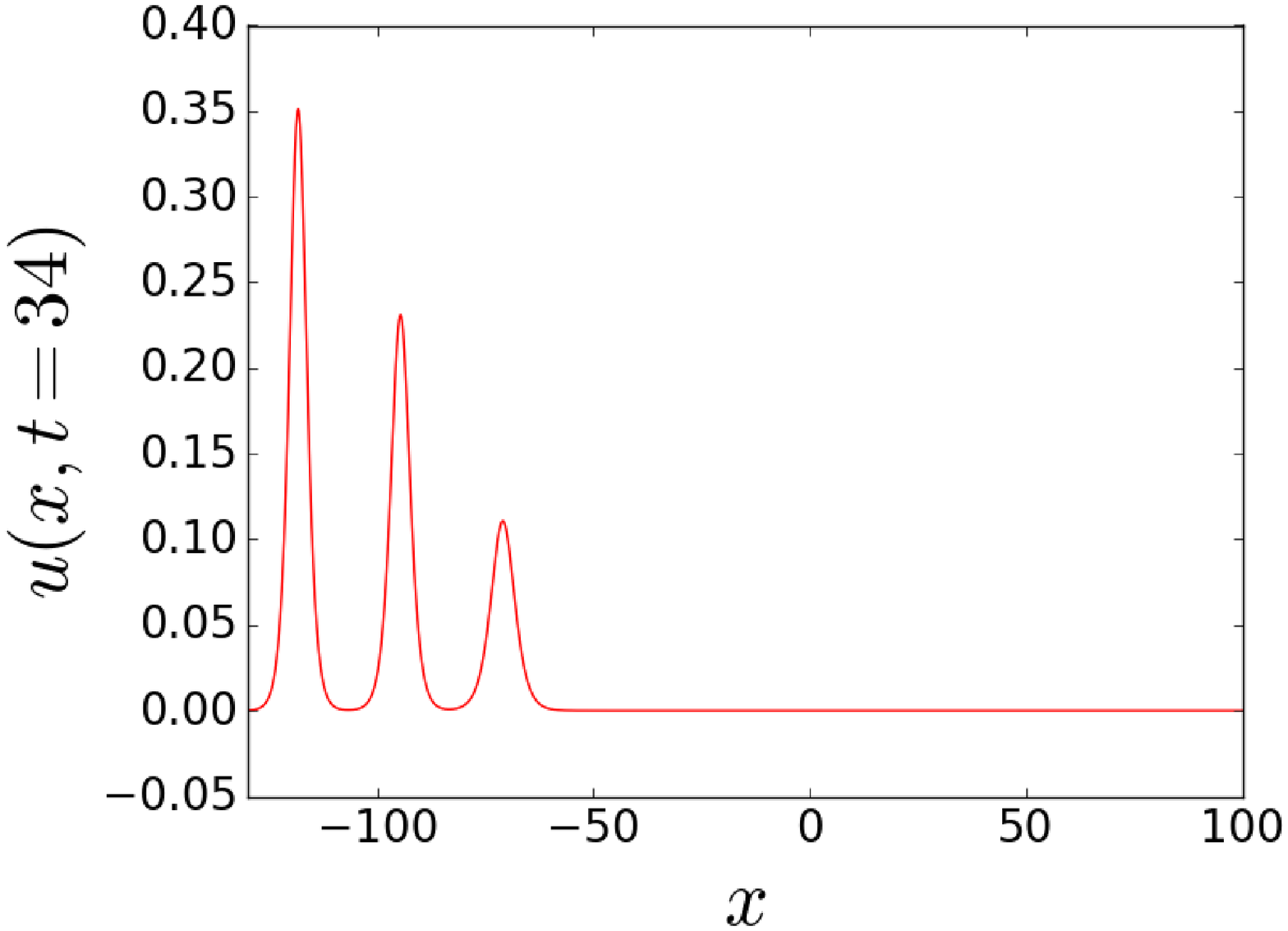}
			\caption{At~$t=34$}
			\label{1t200}
		\end{subfigure}%
		\begin{subfigure}{.34\textwidth}
			\centering
			\includegraphics[scale=0.28]{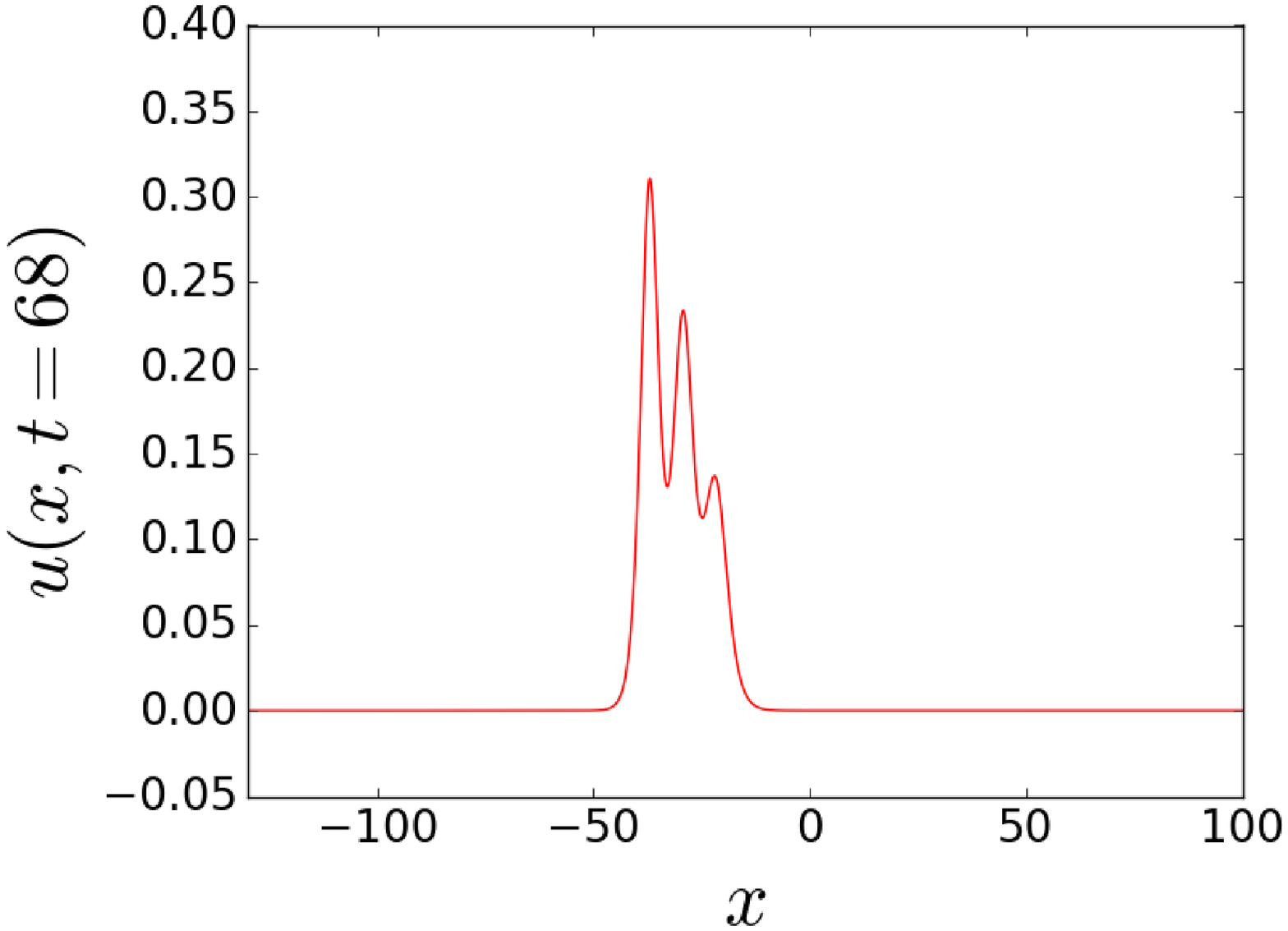}
			\caption{At~$t=68$}
			\label{1t400}
		\end{subfigure}%
		\begin{subfigure}{.34\textwidth}
			\centering
			\includegraphics[scale=0.28]{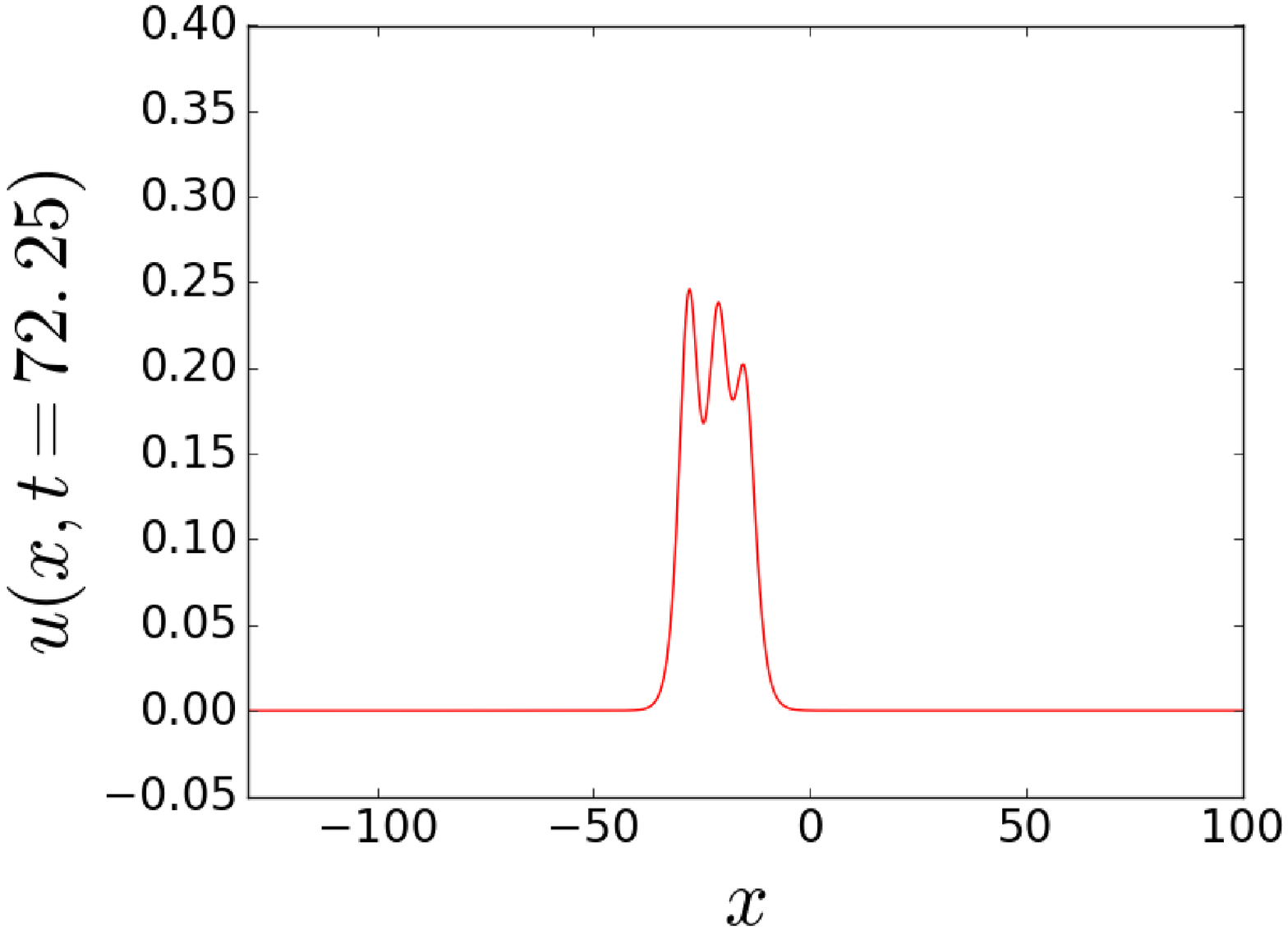}
			\caption{At~$t=72.25$}
			\label{1t425}
		\end{subfigure}
		
		\centering
		\hspace*{-0.1cm}
		\begin{subfigure}{.34\textwidth}
			\centering
			\includegraphics[scale=0.28]{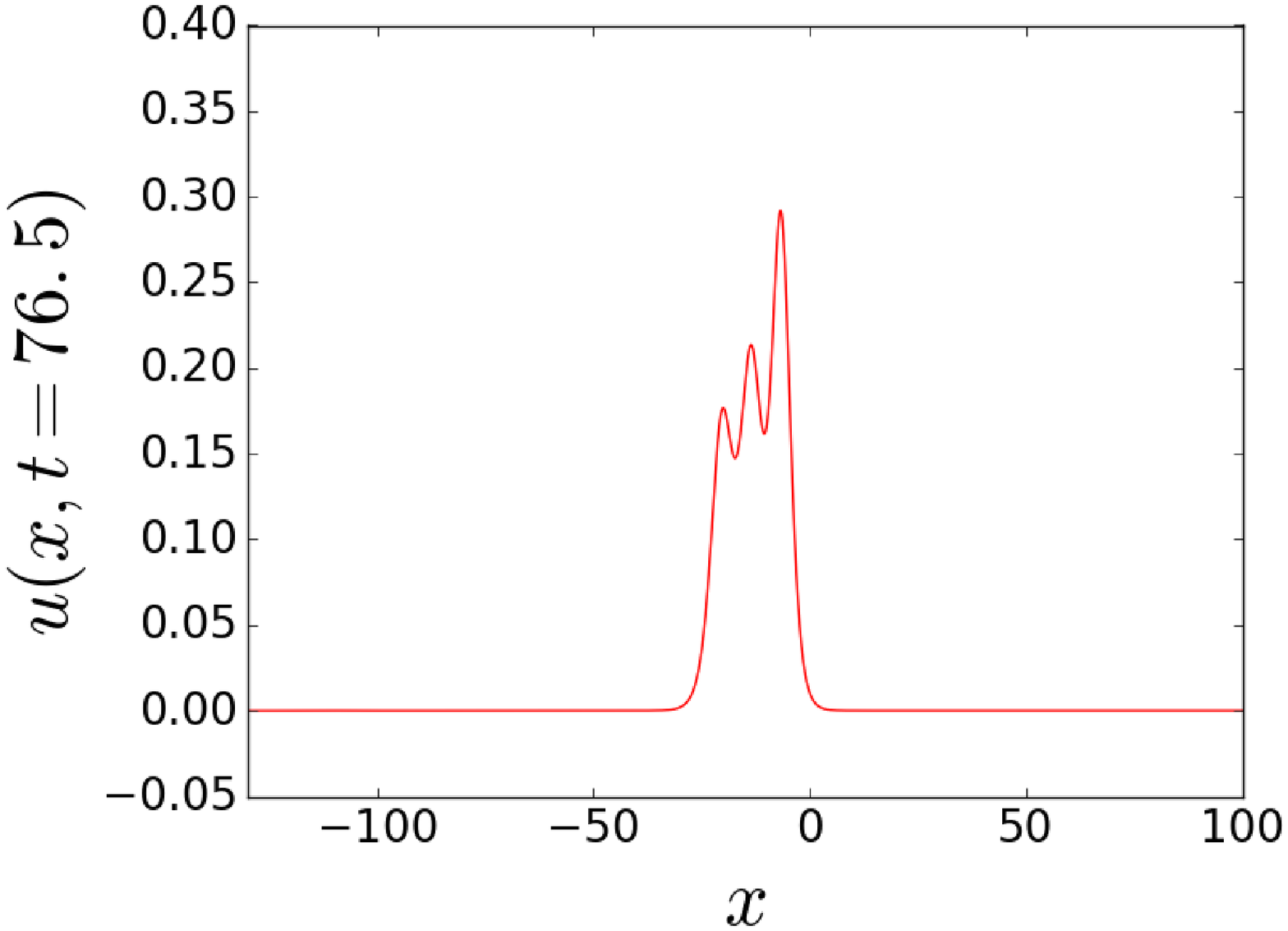}
			\caption{At~$t=76.5$}
			\label{1t450}
		\end{subfigure}%
		\begin{subfigure}{.34\textwidth}
			\centering
			\includegraphics[scale=0.28]{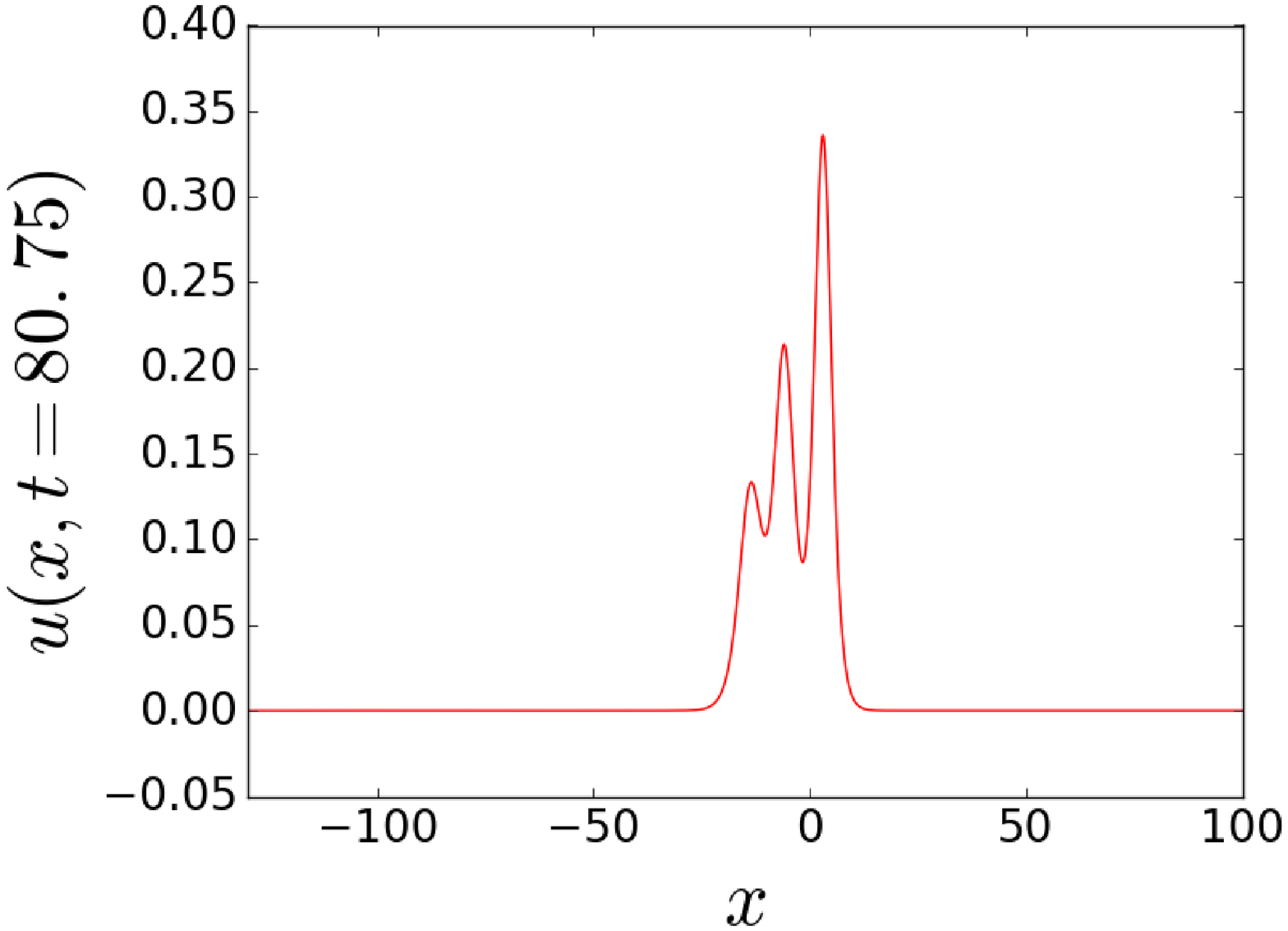}
			\caption{At~$t=80.75$}
			\label{1t475}
		\end{subfigure}%
		\begin{subfigure}{.34\textwidth}
			\centering
			\includegraphics[scale=0.28]{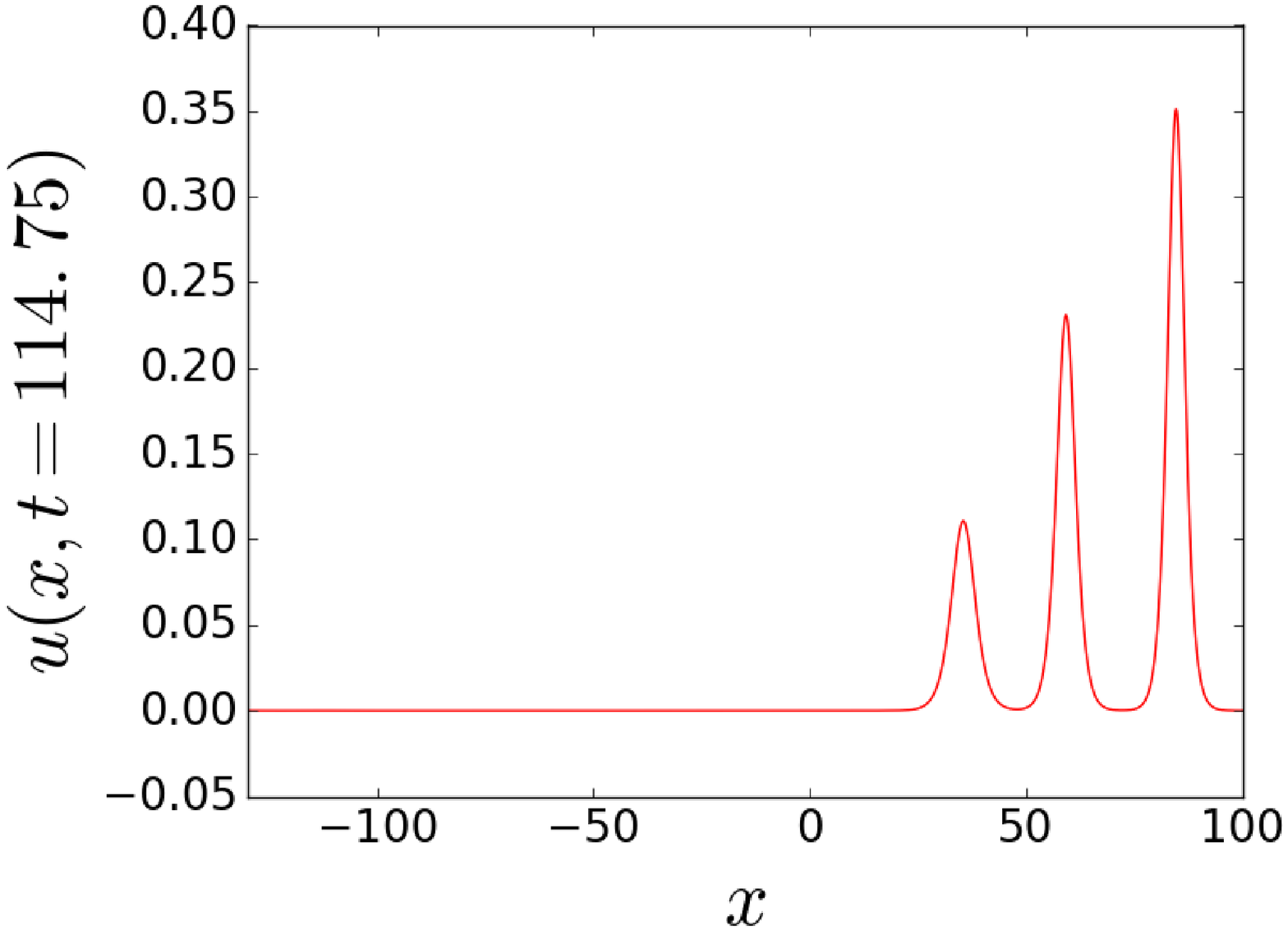}
			\caption{At~$t=114.75$}
			\label{1t675}
		\end{subfigure}
		\caption{This figure shows three solitons interacting with each other at various points in time. Note that at all times during the interaction, there are three distinct maxima present.}
		\label{plot1t200to1t675}
	\end{figure}

\begin{figure}[t!]
		\centering
		\hspace*{-0.1cm}
		\begin{subfigure}{.34\textwidth}
			\centering
			\includegraphics[scale=0.28]{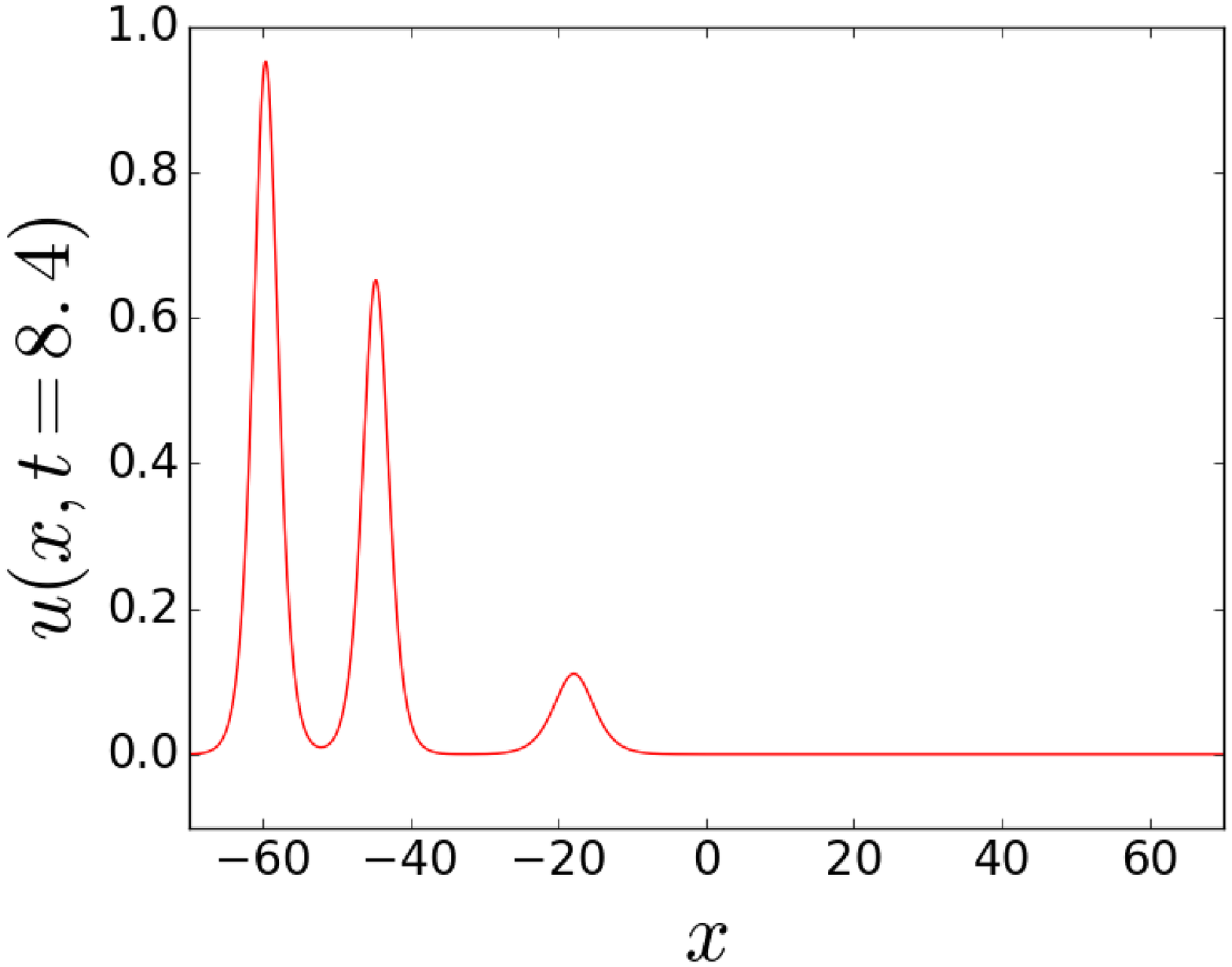}
			\caption{At~$t=8.4$}
			\label{3t200}
		\end{subfigure}%
		\begin{subfigure}{.34\textwidth}
			\centering
			\includegraphics[scale=0.28]{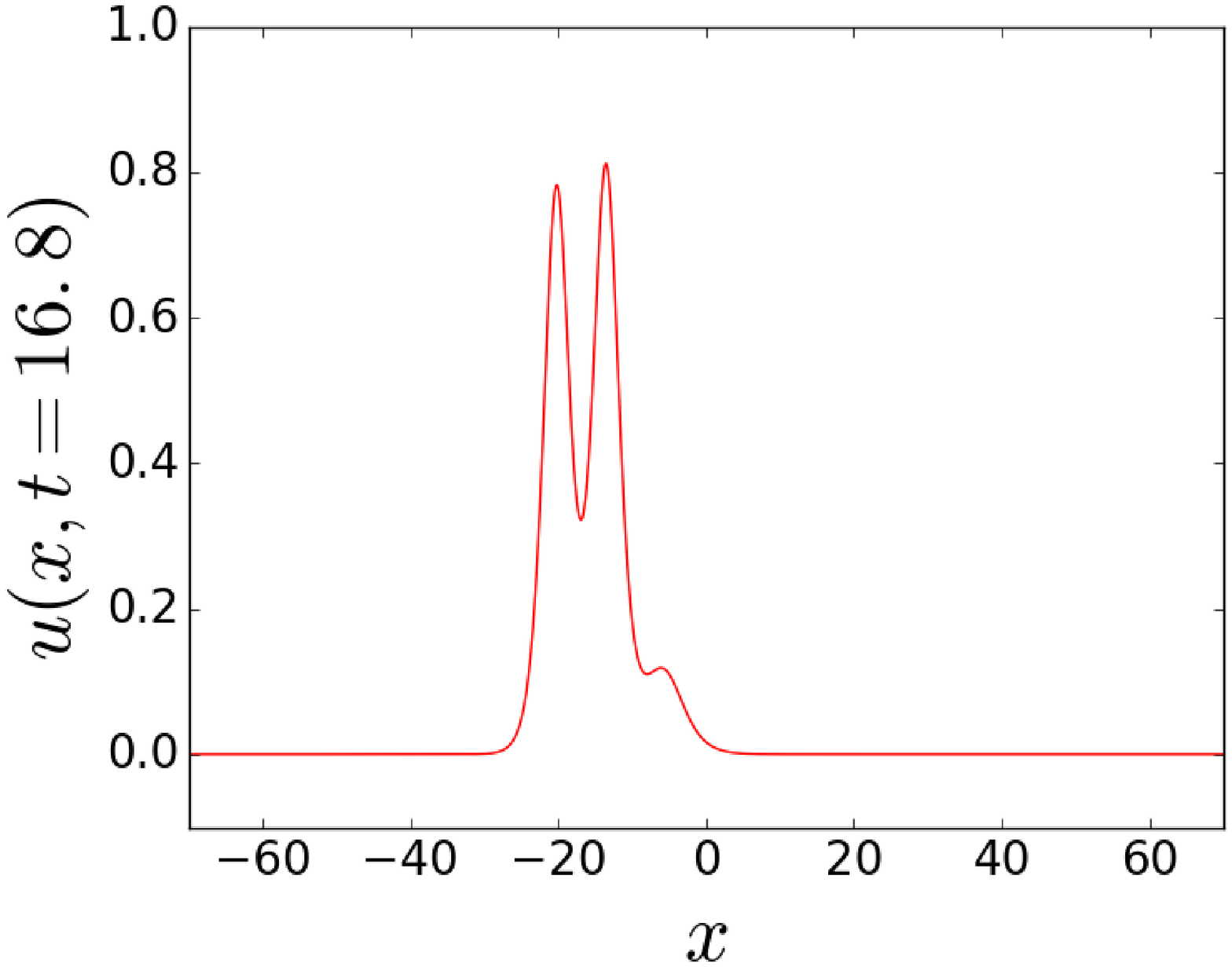}
			\caption{At~$t=16.8$}
			\label{3t400}
		\end{subfigure}%
		\begin{subfigure}{.34\textwidth}
			\centering
			\includegraphics[scale=0.28]{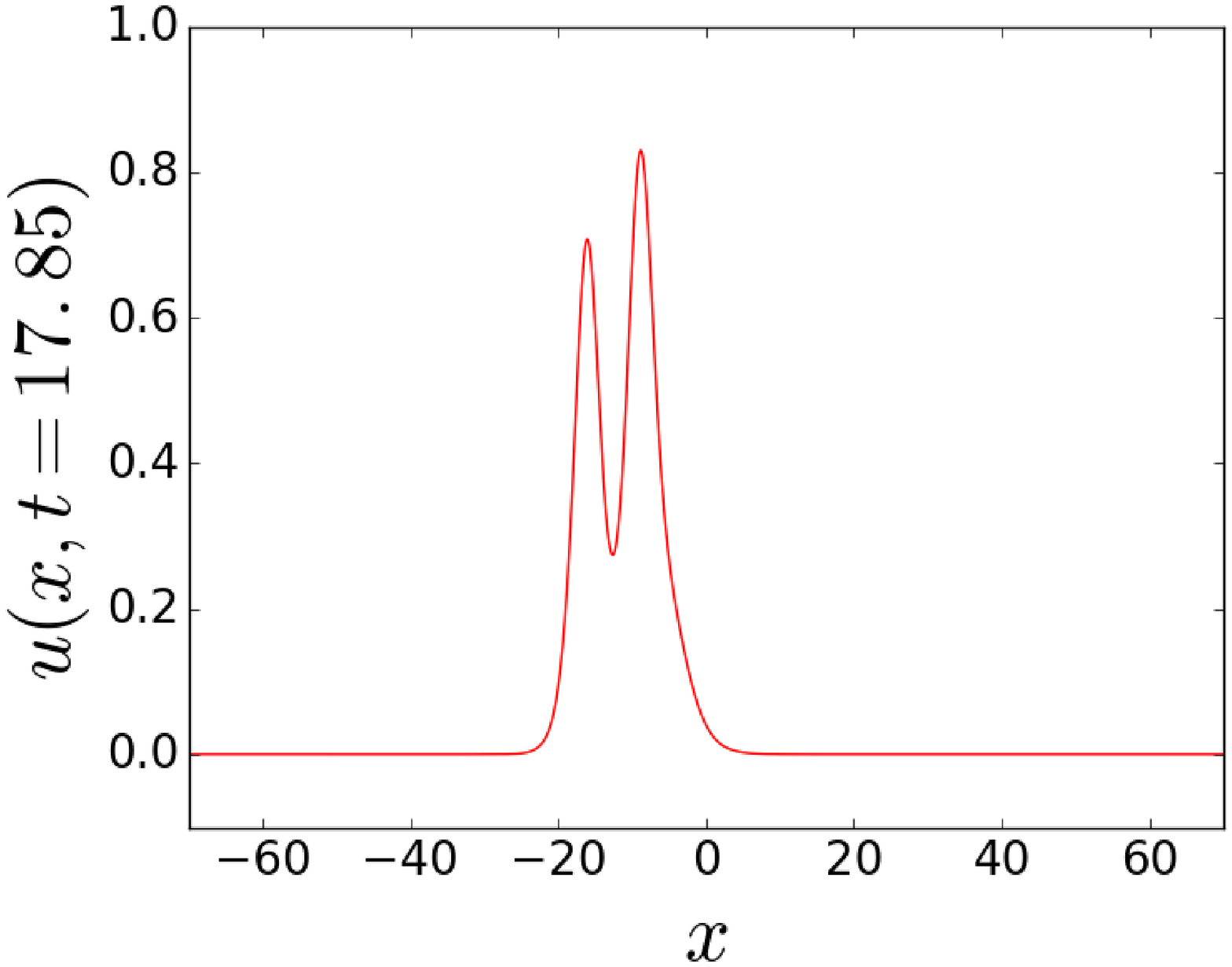}
			\caption{At~$t=17.85$}
			\label{3t425}
		\end{subfigure}
		
		\centering
		\hspace*{-0.1cm}
		\begin{subfigure}{.34\textwidth}
			\centering
			\includegraphics[scale=0.28]{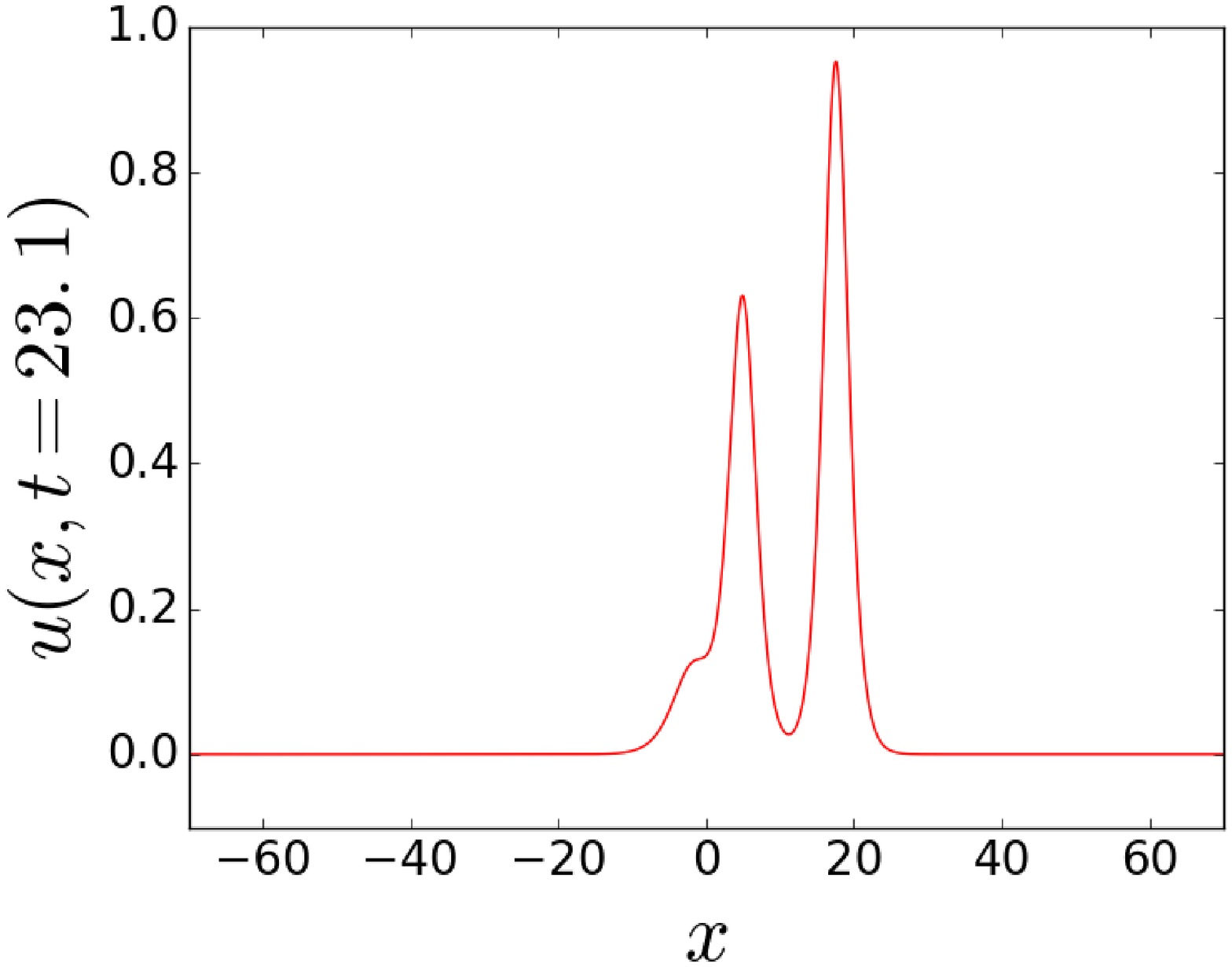}
			\caption{At~$t=23.1$}
			\label{3t550}
		\end{subfigure}%
		\begin{subfigure}{.34\textwidth}
			\centering
			\includegraphics[scale=0.28]{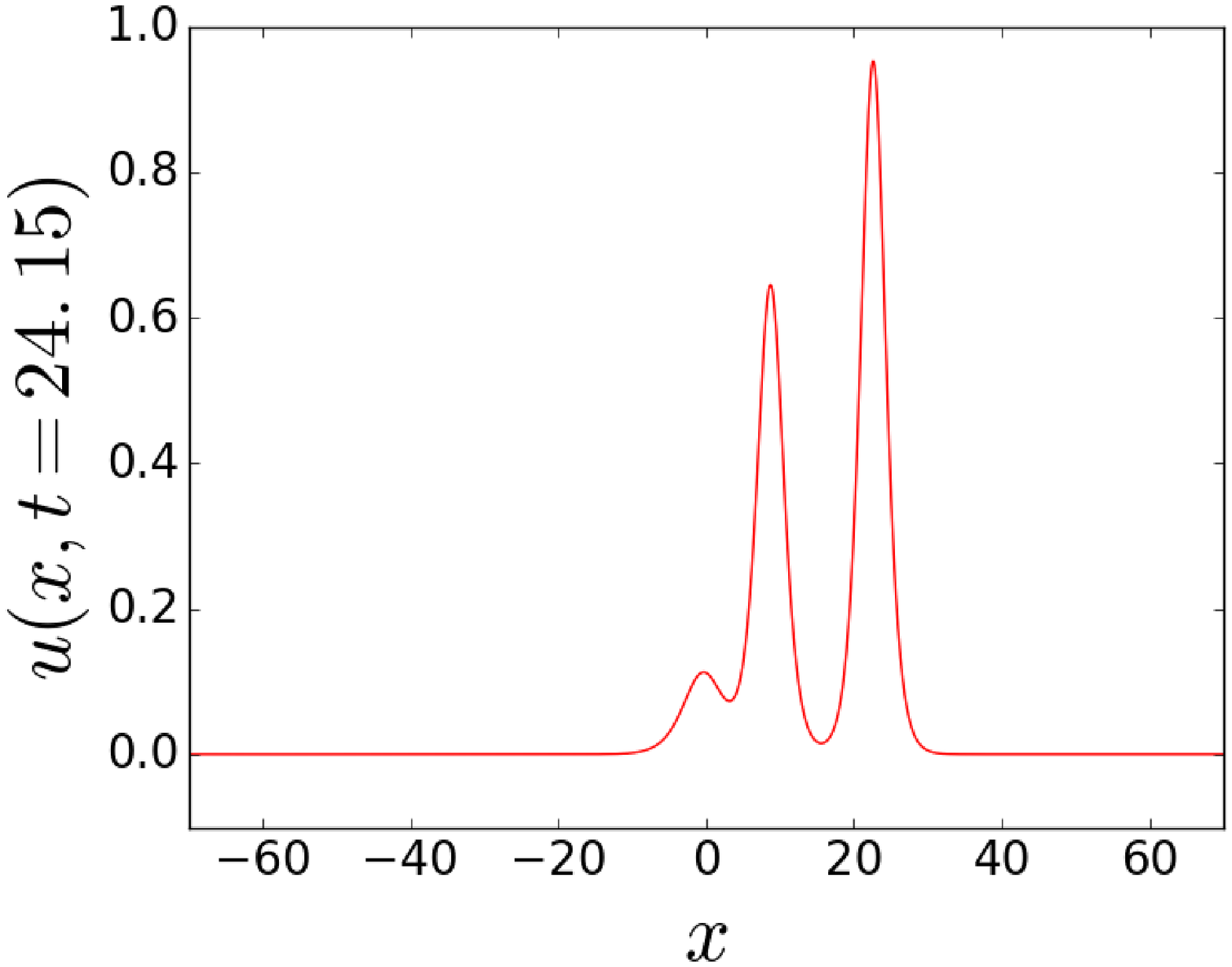}
			\caption{At~$t=24.15$}
			\label{3t575}
		\end{subfigure}%
		\begin{subfigure}{.34\textwidth}
			\centering
			\includegraphics[scale=0.28]{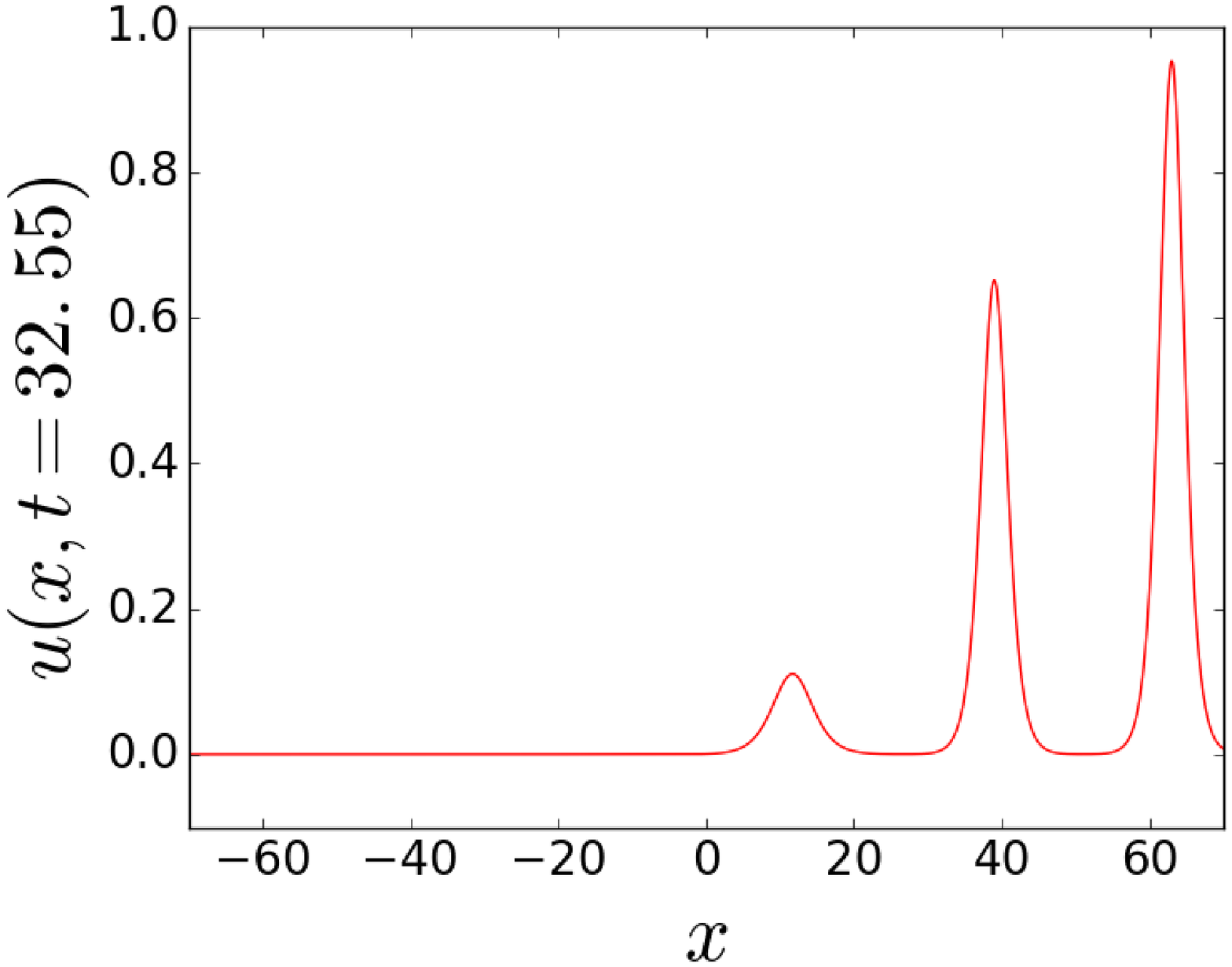}
			\caption{At~$t=32.55$}
			\label{3t775}
		\end{subfigure}
		\caption{This figure shows three solitons interacting with each other at various points in time. Note that there is a small period of time ($17 \lesssim t \lesssim 23$) during the collision where the smallest wave is `absorbed' and there are only two distinct maxima present (see figure~\ref{3t425} for example).} 
		\label{plot3t200to1t775}
	\end{figure}

		
	We are primarily interested in three-soliton interactions,
 and so for the simulations discussed in this section the phase constants $\delta_i$ have  always been chosen in
 such a way that all three solitons scatter with each other at more or less the same time. 
To illustrate this, figures~\ref{plot1t200to1t675}
	and~\ref{plot3t200to1t775} show two of such interactions, where both simulations start at $t=0$ with 
the solitons placed far apart of each other. 
	
\begin{figure}[b!]
		\centering
		\hspace*{-0.1cm}
		\begin{subfigure}{.34\textwidth}
			\centering
			\includegraphics[scale=0.28]{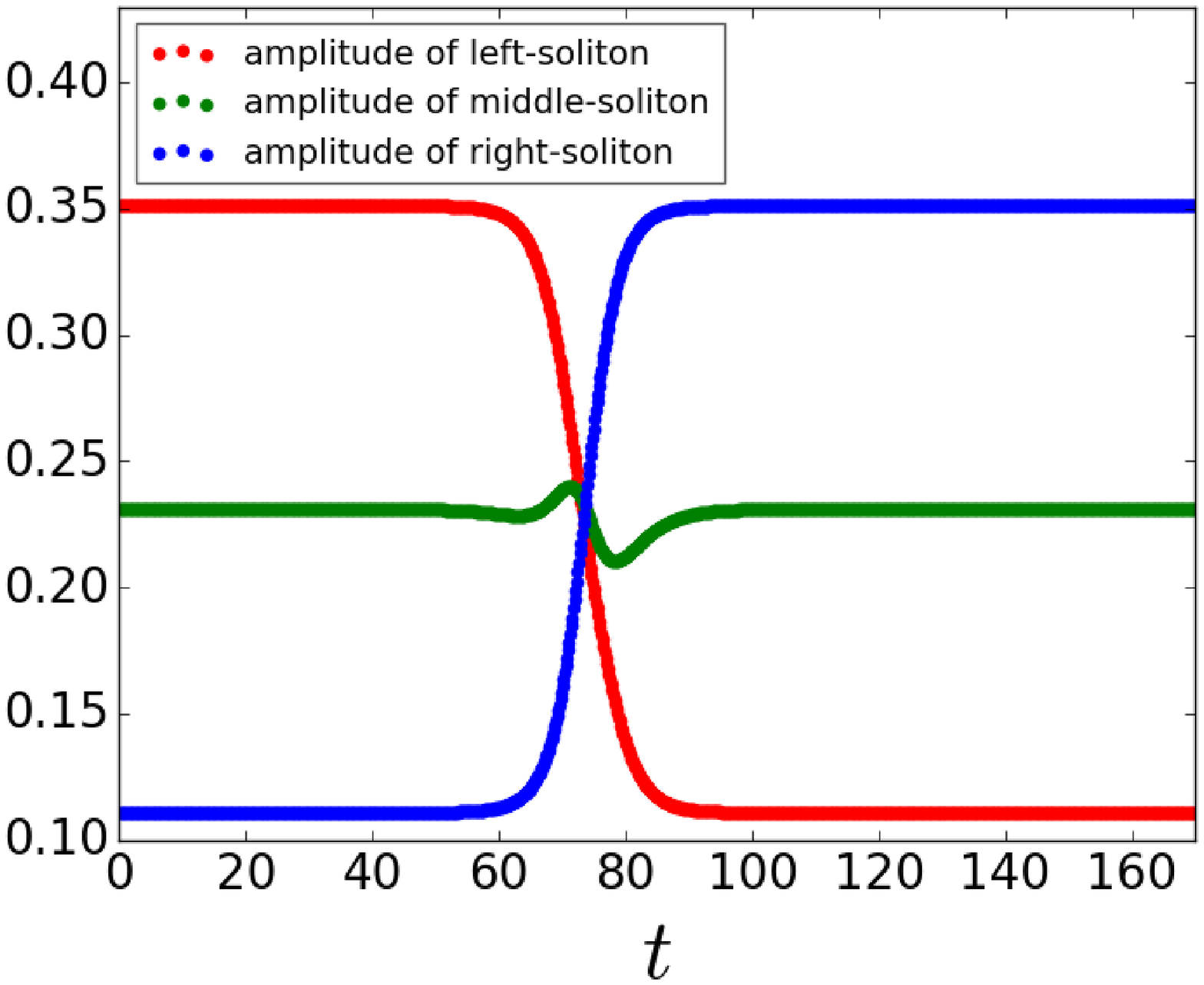}
			\caption{}
			\label{1amplitude_vs_time}
		\end{subfigure}%
		\begin{subfigure}{.34\textwidth}
			\centering
			\includegraphics[scale=0.28]{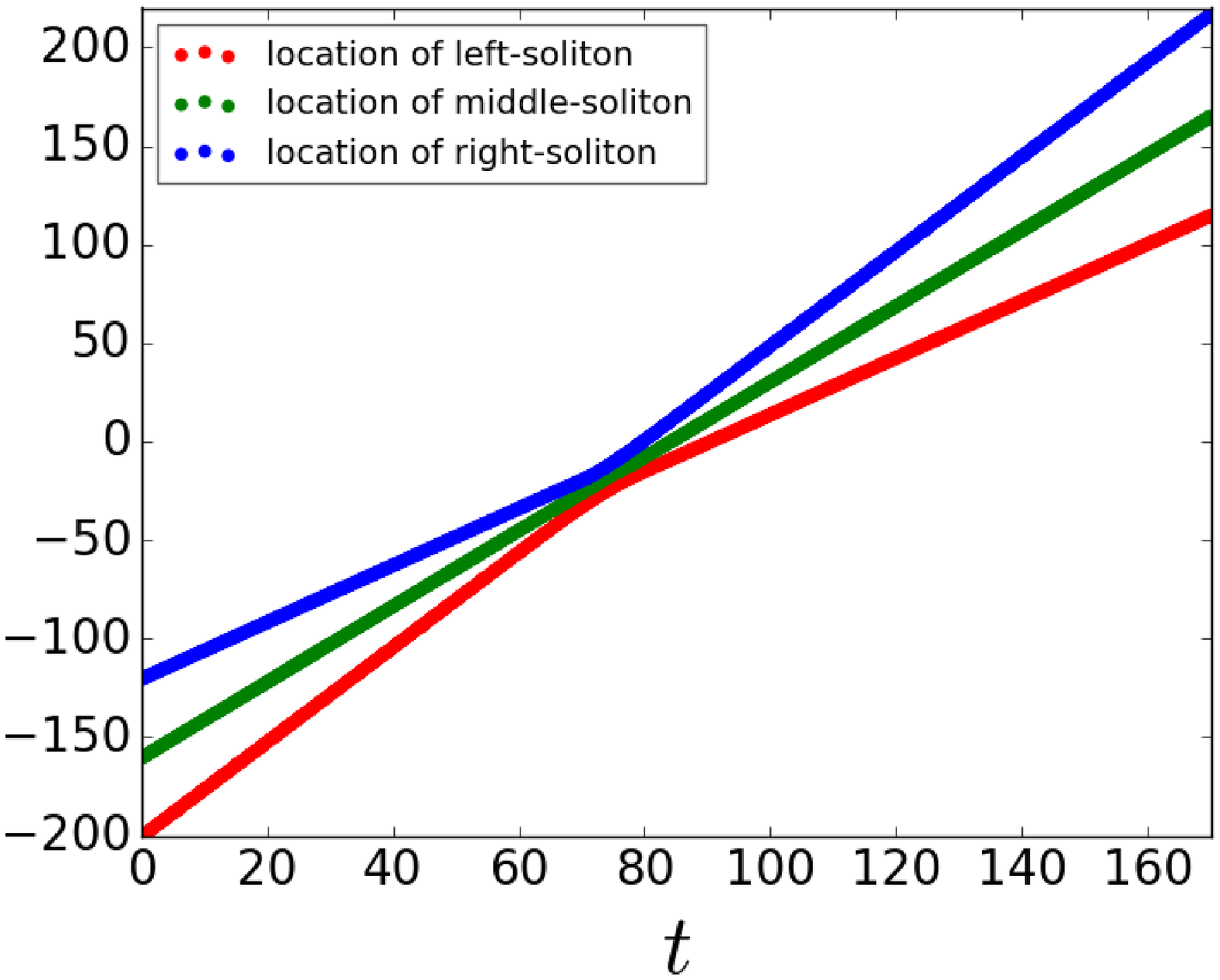}
			\caption{}
			\label{1location_vs_time}
		\end{subfigure}%
		\begin{subfigure}{.34\textwidth}
			\centering
			\includegraphics[scale=0.28]{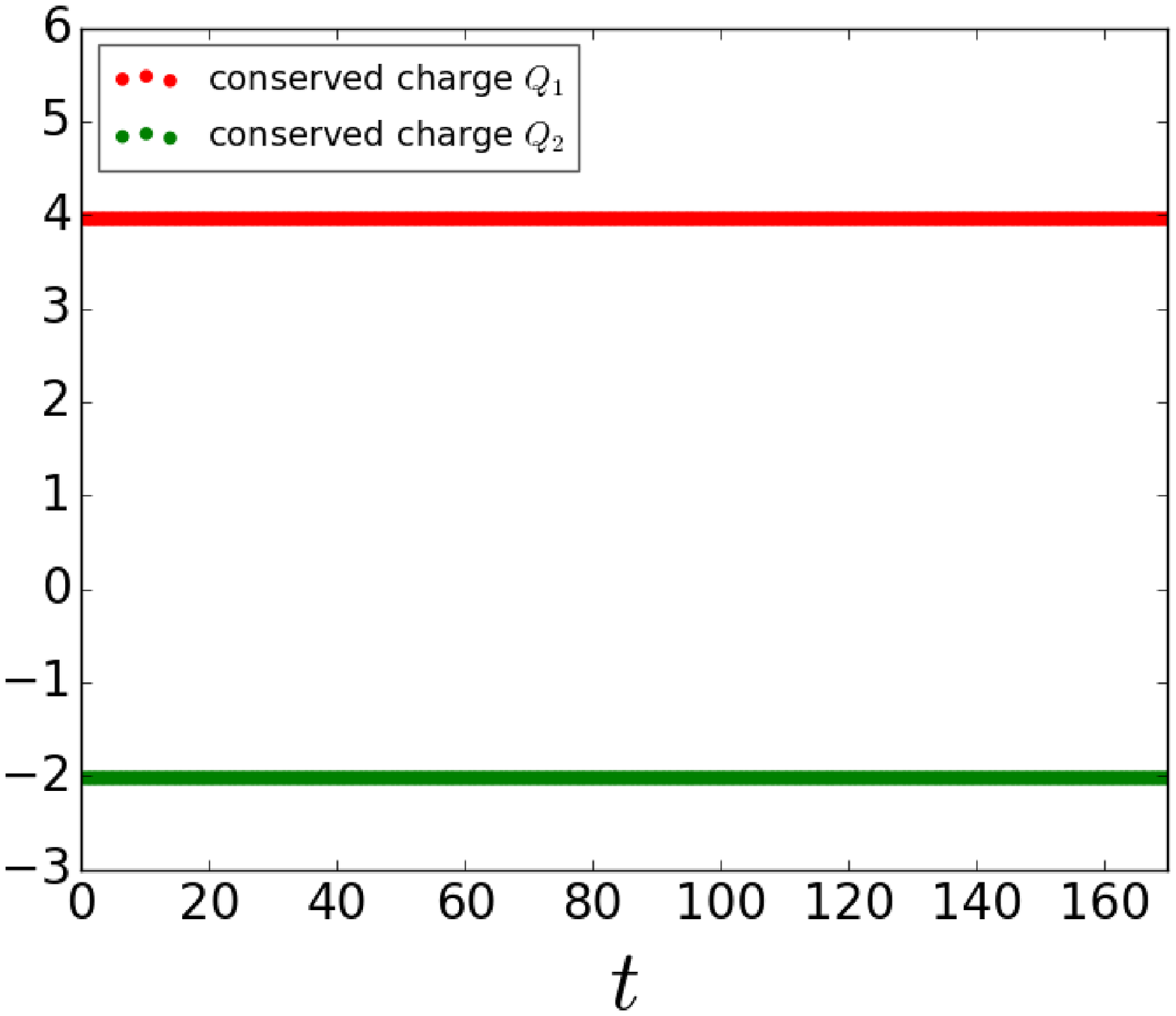}
			\caption{}
			\label{1charge_vs_time}
		\end{subfigure}
		\caption{For figures~\ref{1amplitude_vs_time} and~\ref{1location_vs_time}, the red dots present the time dependence of the amplitude and the location of the left-soliton of the simulation shown in figure~\ref{plot1t200to1t675}. Similarly, the green and blue dots correspond to the middle- and right-soliton, respectively. Finally, figure~\ref{1charge_vs_time} shows how the conserved charges of the corresponding simulation vary with time.}
		\label{1amplitude_vs_time_and_1location_vs_time}
	\end{figure}

\begin{figure}[b!]
		\centering
		\hspace*{-0.1cm}
		\begin{subfigure}{.34\textwidth}
			\centering
			\includegraphics[scale=0.28]{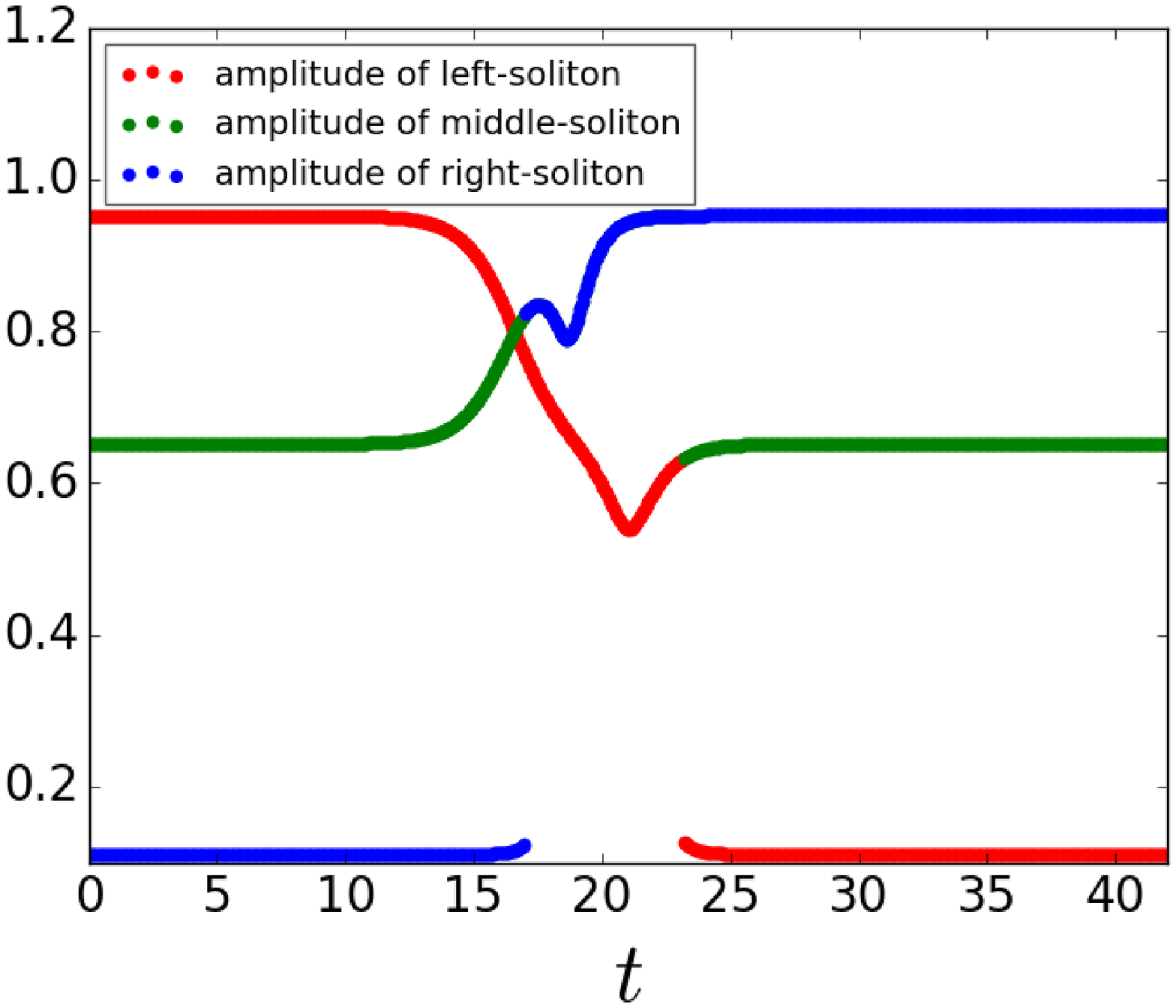}
			\caption{}
			\label{3amplitude_vs_time}
		\end{subfigure}%
		\begin{subfigure}{.34\textwidth}
			\centering
			\includegraphics[scale=0.28]{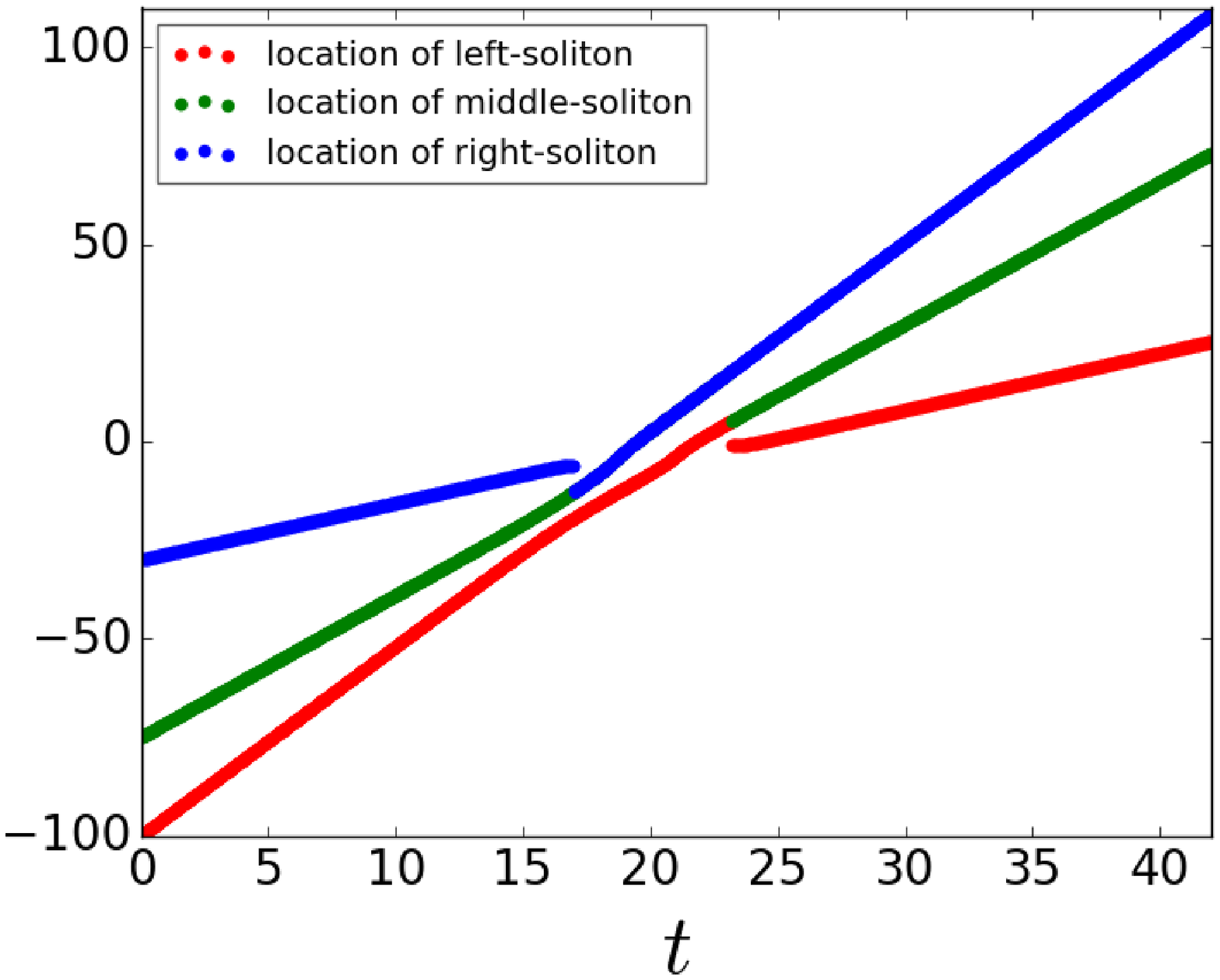}
			\caption{}
			\label{3location_vs_time}
		\end{subfigure}%
		\begin{subfigure}{.34\textwidth}
			\centering
			\includegraphics[scale=0.28]{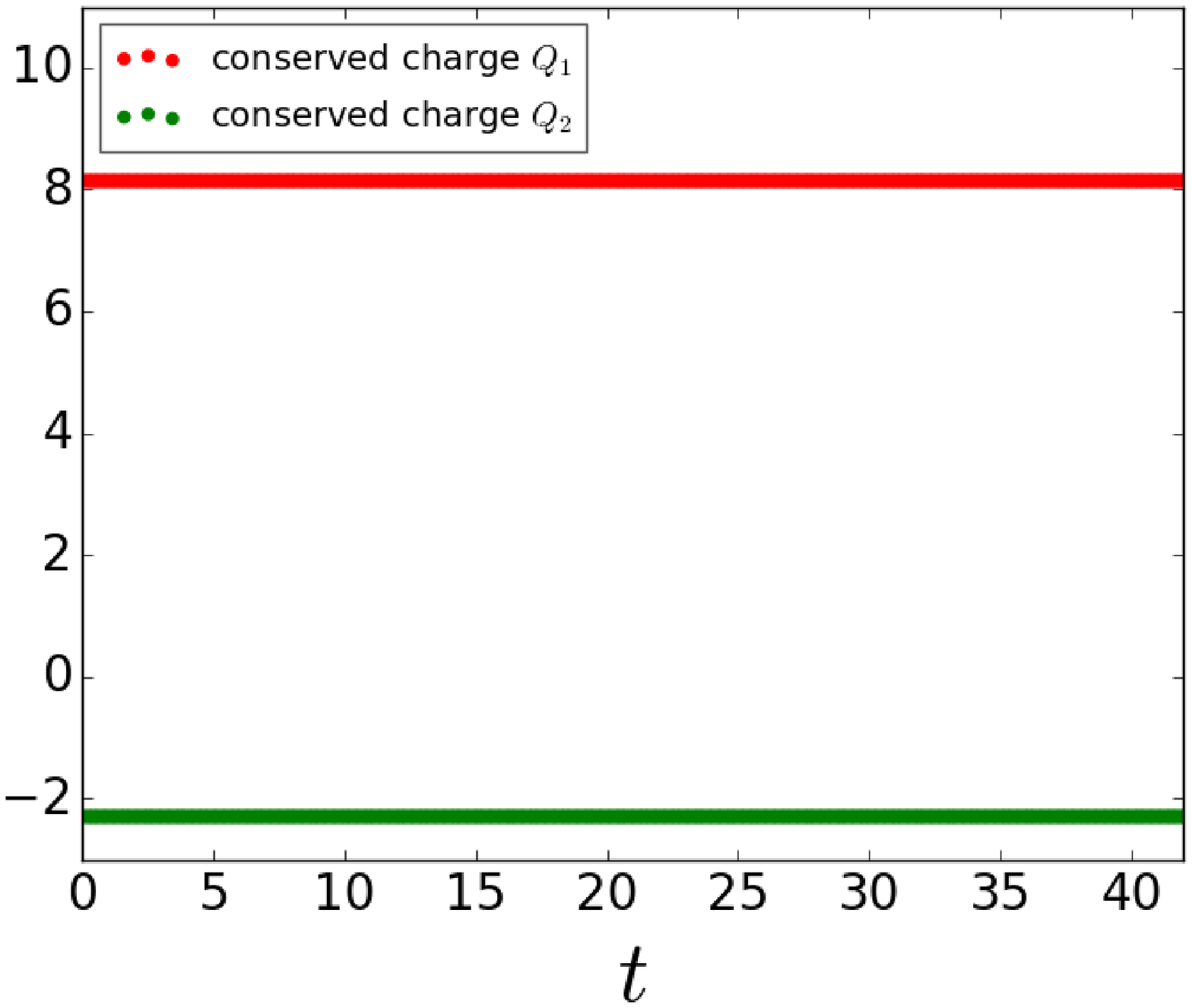}
			\caption{}
			\label{3charge_vs_time}
		\end{subfigure}
		\caption{For figures~\ref{3amplitude_vs_time} and~\ref{3location_vs_time}, the red dots present the time dependence of the amplitude and the location of the left-soliton of the simulation shown in figure~\ref{plot3t200to1t775}. Similarly, the green and blue dots correspond to the middle- and right-soliton, respectively. Finally, figure~\ref{3charge_vs_time} shows how the conserved charges of the corresponding simulation vary with time.}
		\label{3amplitude_vs_time_and_1location_vs_time}
	\end{figure}

For these two simulations, figures~\ref{1amplitude_vs_time_and_1location_vs_time}
	and~\ref{3amplitude_vs_time_and_1location_vs_time}
	show, respectively,  the time dependence of the amplitude and the location 
(defined as the location of the amplitude of the wave) of each soliton, and the time dependence 
of the charges~$Q_1$ and~$Q_2$. Using these results, we have observed that for all our simulations, after 
the three-soliton interaction, the solitons recovered their original amplitudes and velocities with a
 numerical error of less than~$0.1 \%$. Furthermore, the quantities~$Q_1$ and~$Q_2$ have been incredibly
 well conserved for all our simulations with a numerical error of less than~$0.001 \%$ compared 
to the approximations given by equations~(\ref{1.5}) and~(\ref{1.6}). These results 
demonstrate that the solitons, in these configurations, behave as solitons in integrable models.

	Assuming that the total phase-shift each soliton experiences when three solitons collide
 is the sum of the pairwise phase-shifts, then, analytically, we expect the location of the largest soliton
 (related to~$\eta_1$) to experience a phase-shift forward by~$\ln(A_{12})+\ln(A_{13})$,
 the soliton corresponding to~$\eta_2$ to experience a phase-shift given by~$-\ln(A_{12})+\ln(A_{23})$ 
and the smallest soliton to be phase-shifted by~$-\ln(A_{13})-\ln(A_{23})$, where~$A_{13}$ and~$A_{23}$
 are defined in a similar way as~$A_{12}$ (see equation~(\ref{1.2.1})). We found that the phase-shifts
 for all our simulations have always been within~$5.0\%$ of the expected analytic value.
 Since these errors are consistent with the errors discussed near the end of
 section~\ref{Modified_regularized_long-wave_equation_equation}, we conclude that the phase-shifts
 experienced by each soliton during a numerical three-soliton interaction is additive, which 
suggest the non-existence of additional `three-body' forces, ({\it i.e.}, the phase shifts can
 be explained by the additivity of `two-body' forces). This is another indicator that the solitons
 of our model behave like solitons in an integrable model.

	\section{Soliton resolution conjecture} \label{soliton_resolution_conjecture}
	
	In this section we investigate the time evolution of soliton-like lumps which are not
 analytical solutions of the mRLW equation.
 This is a test of the idea that such initial field configurations may eventually decouple 
into soliton-like and radiation-like components. For many non-linear dispersive equations there
 is evidence suggesting that such arbitrary finite-energy initial configurations always decouple in 
such a manner and this expectation is also referred to as the soliton resolution
 conjecture~\cite{tao}.\footnote{We thank A. Hone for drawing our attention to this conjecture.}
	
	Of course the choice of the initial lump-like configuration is very arbitrary -
 one could take a Gaussian field or any other more complicated initial condition field
for which the $u=-q_{xt}$ resembles a soliton-like structure. However, since our numerical scheme
 requires an analytical expression for~$q$ and~$q_t$ as initial conditions, and the field we 
are interested in is described by~$u=-q_{xt}$, we are somewhat limited in our choices. 
	
\begin{figure}[b!]
		\centering
		\hspace*{-0.1cm}
		\begin{subfigure}{.34\textwidth}
			\centering
			\includegraphics[scale=0.28]{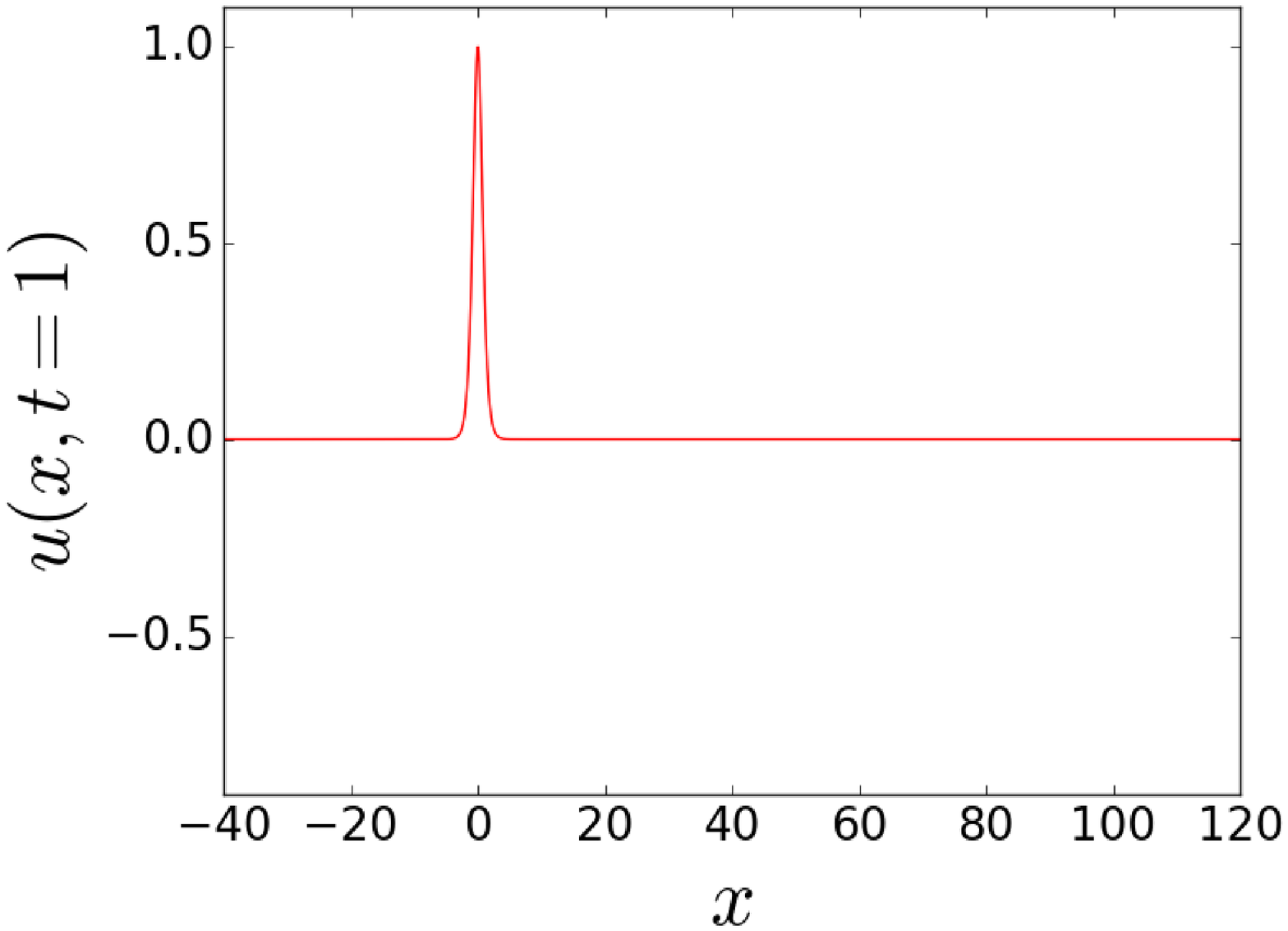}
			\caption{At~$t=1$}
			\label{8_000}
		\end{subfigure}%
		\begin{subfigure}{.34\textwidth}
			\centering
			\includegraphics[scale=0.28]{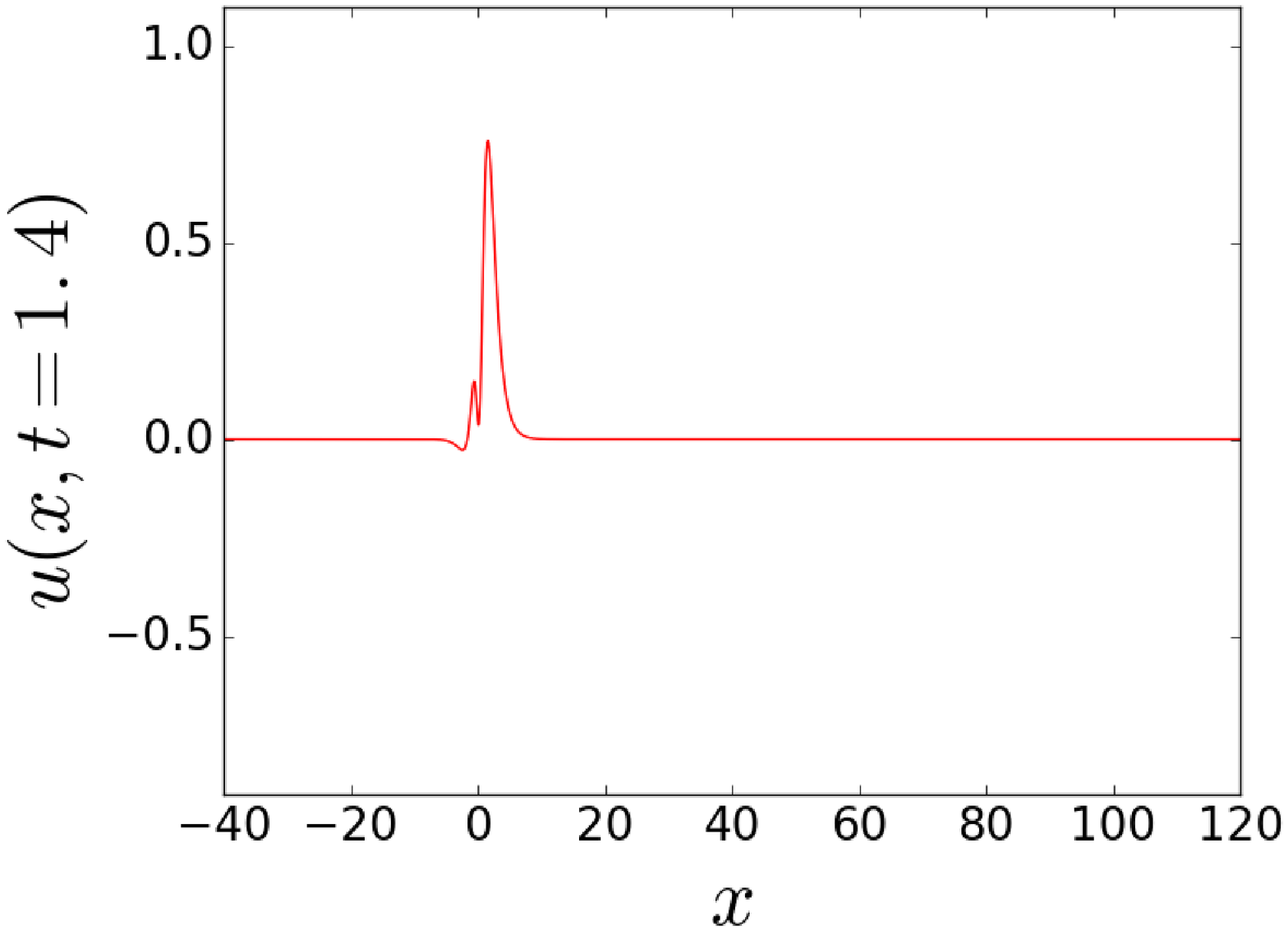}
			\caption{At~$t=1.4$}
			\label{8_050}
		\end{subfigure}%
		\begin{subfigure}{.34\textwidth}
			\centering
			\includegraphics[scale=0.28]{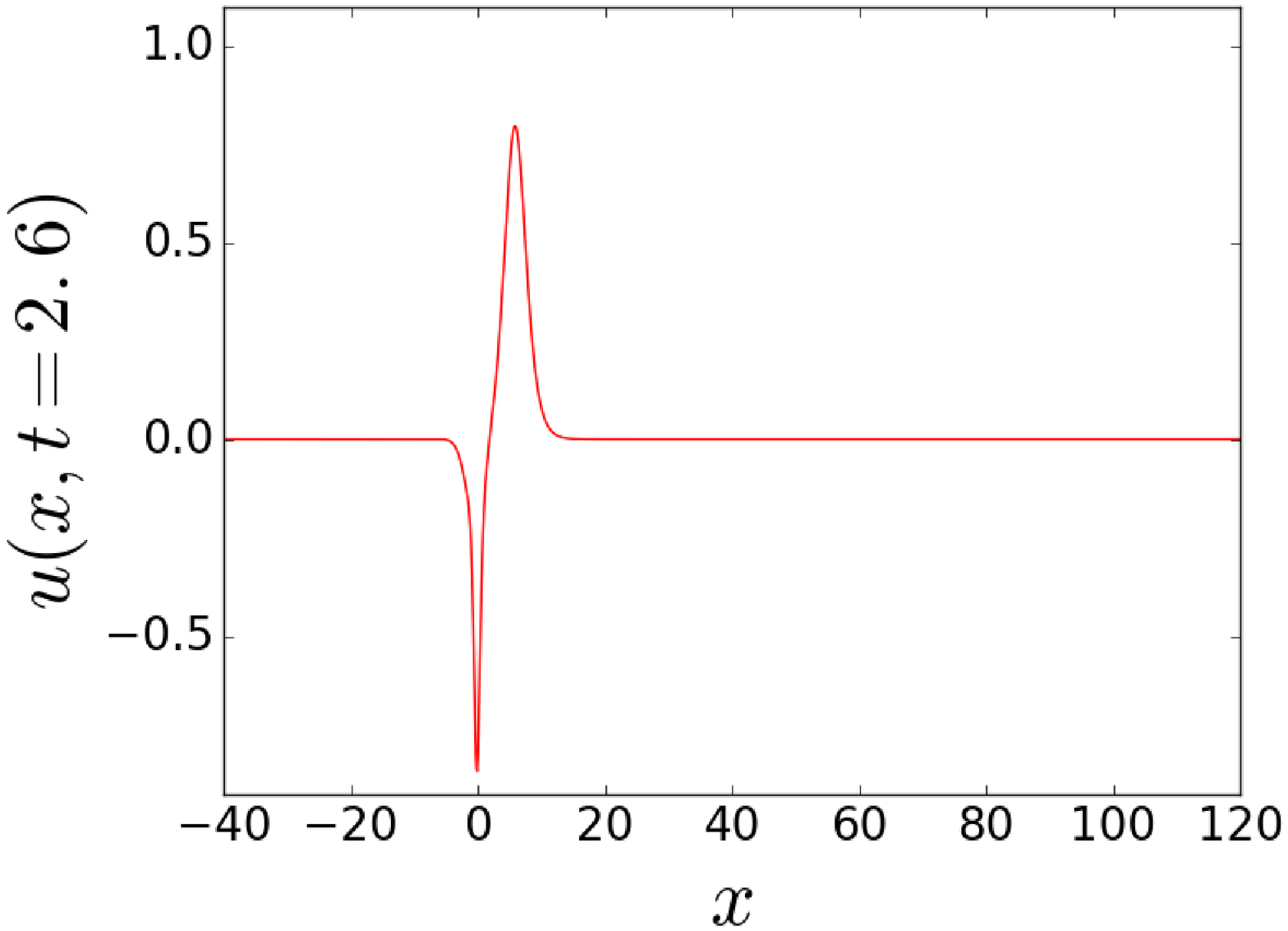}
			\caption{At~$t=2.6$}
			\label{8_100}
		\end{subfigure}
			
		\centering
		\hspace*{-0.1cm}
		\begin{subfigure}{.34\textwidth}
			\centering
			\includegraphics[scale=0.28]{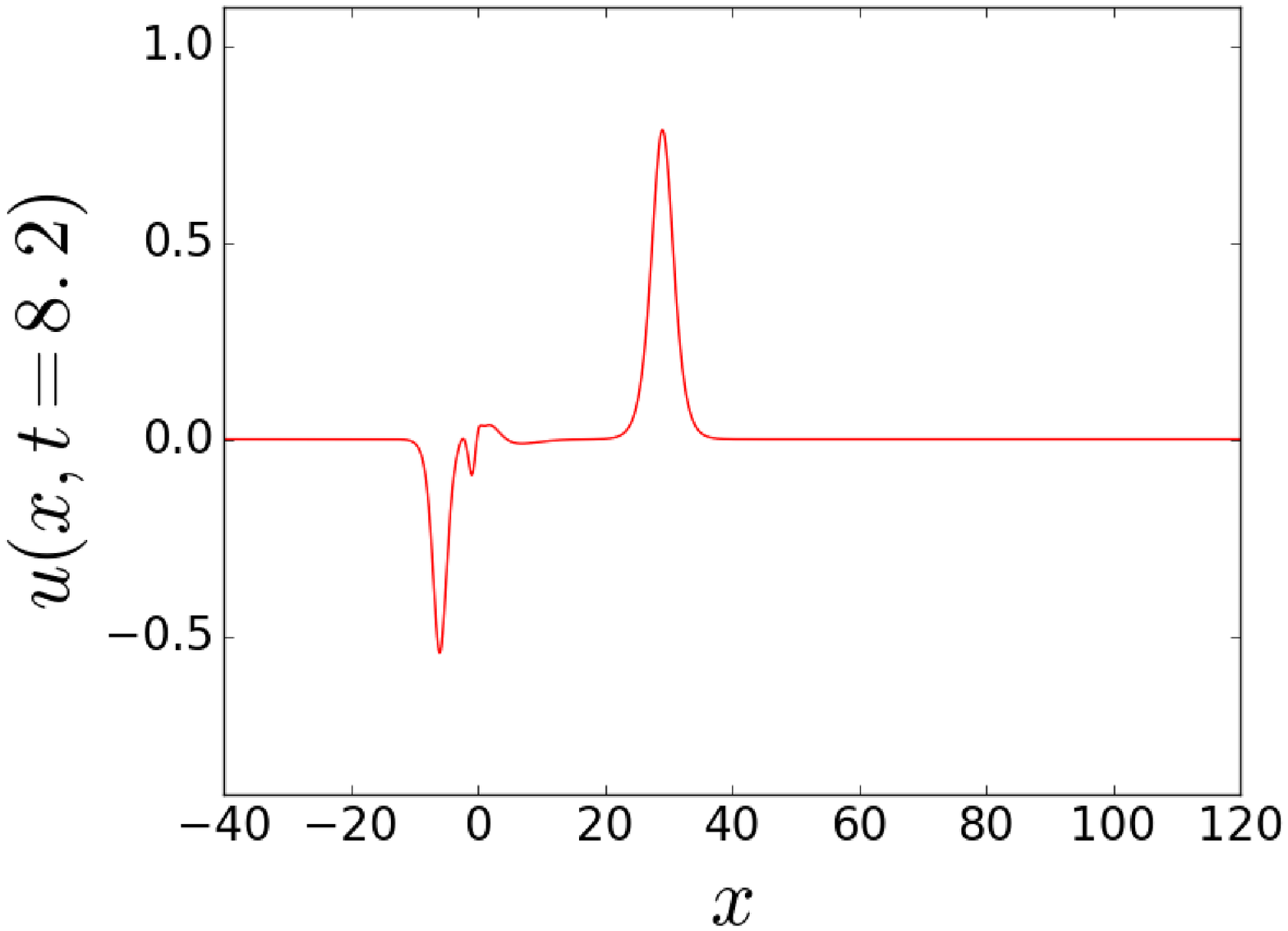}
			\caption{At~$t=8.2$}
			\label{8_150}
		\end{subfigure}%
		\begin{subfigure}{.34\textwidth}
			\centering
			\includegraphics[scale=0.28]{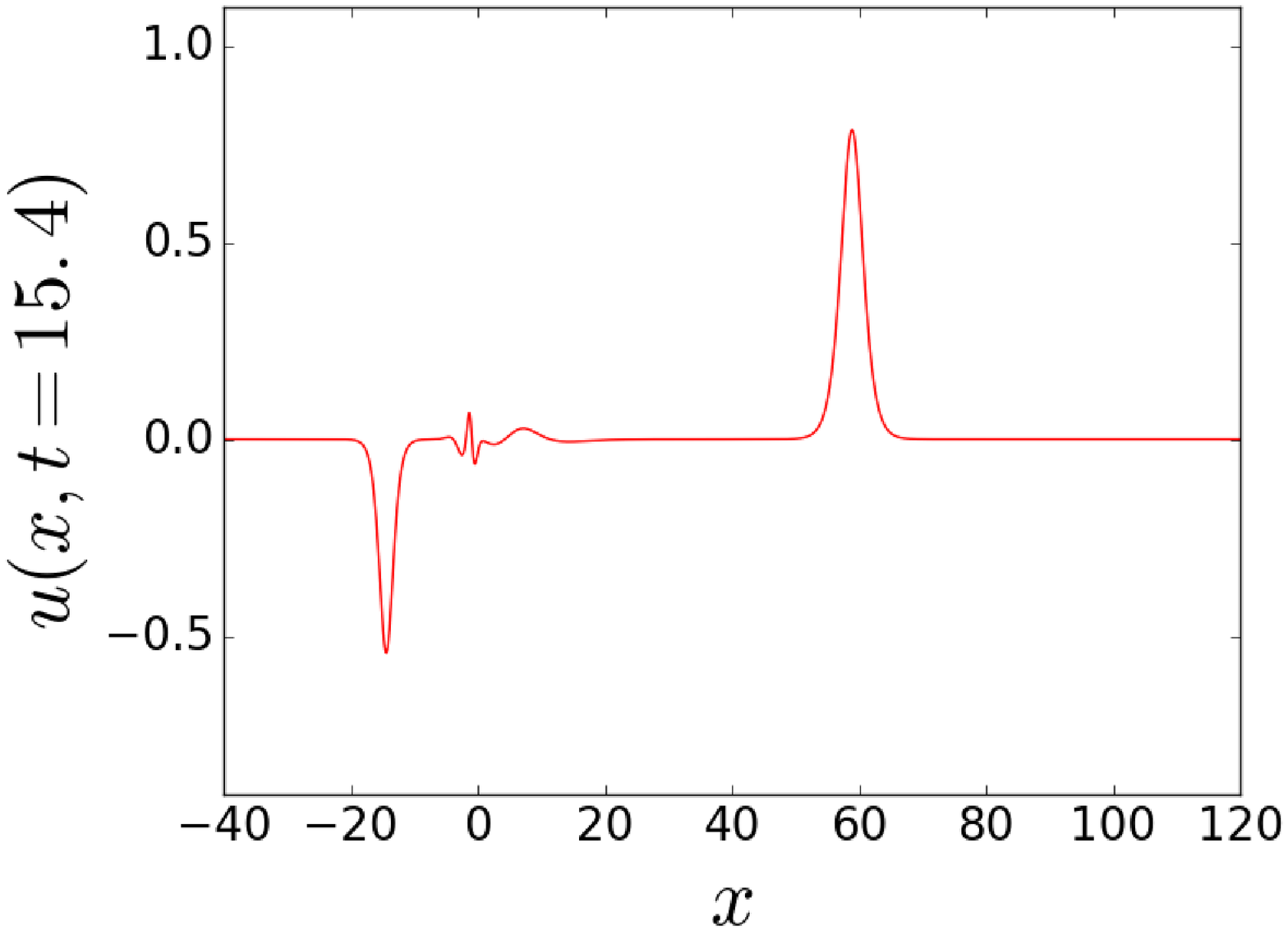}
			\caption{At~$t=15.4$}
			\label{8_200}
		\end{subfigure}%
		\begin{subfigure}{.34\textwidth}
			\centering
			\includegraphics[scale=0.28]{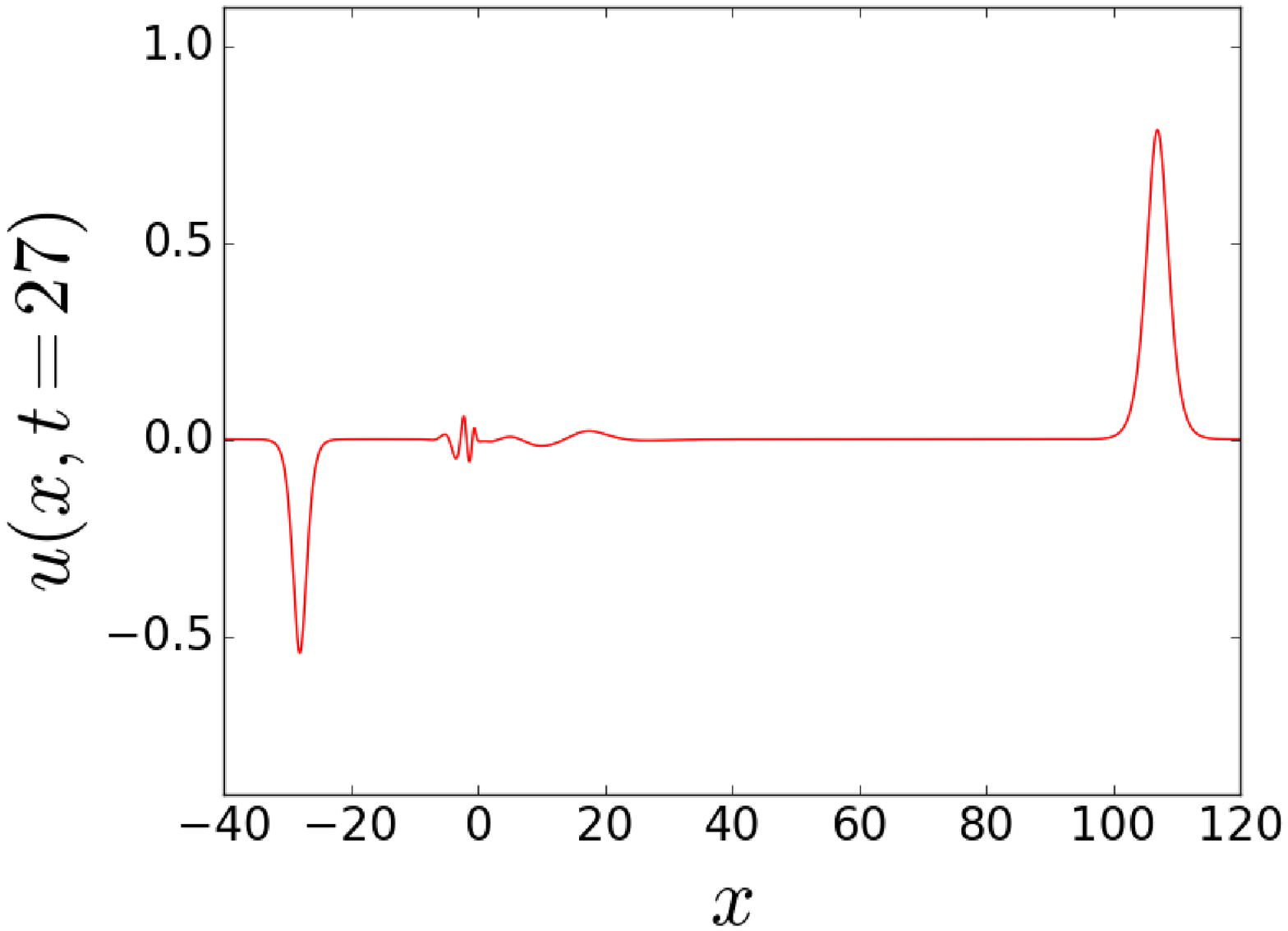}
			\caption{At~$t=27$}
			\label{8_250}
		\end{subfigure}
		\caption{Plots of $x$-dependence of $u(x,t)$ (at various values of $t$) of the numerical evolution of $q$ field
			with the initial condition taken from equation~(\ref{2.2}).}
		\label{plot8_000to6_250}
	\end{figure}

	
	Figure~\ref{plot8_000to6_250}
	shows the time evolution of our first choice of the lump given by
	\begin{equation}
		q = -t \tanh(x) \implies u = \sech^2(x) \,. \label{2.2}
	\end{equation}
	We started this simulation at $t=1$ and, as fig~\ref{8_000} shows, the initial conditions describe a lump with a positive amplitude located at~$x=0$. Then
 figures~\ref{8_050} and~\ref{8_100} show that this initial configuration evolved into a lump and 
an anti-lump configuration. The anti-lump traveled to the left leaving behind some `radiation', whereas
 the positive lump moved to the right without emitting any (visible) `radiation' 
(see figures~\ref{8_150},~\ref{8_200}, and~\ref{8_250}). The `radiation' left behind by the anti-lump was
 also slowly moving to the left and (although it is difficult to see this by looking at 
figure~\ref{plot8_000to6_250}) it also emitted some further `radiation' which started to travel to the right
 (but not fast enough to catch up with the positive lump). We ran this simulation until $t = 241$ without the 
system blowing up. The simulation might blow up if we run it for a longer time, but we lacked the computational
 power for studying this further. Let us add that when we repeated this simulation with a negative amplitude 
({\it i.e.}, using~$u = -\sech^2(x)$ as initial conditions), then the simulation blew up at approximately
 $t = 28.9$. From this we see that at least the `lump-like' described by equation~(\ref{2.2}) was reasonably robust
and long-lived. 
	
\begin{figure}[b!]
		\centering
		\hspace*{-0.1cm}
		\begin{subfigure}{.34\textwidth}
			\centering
			\includegraphics[scale=0.28]{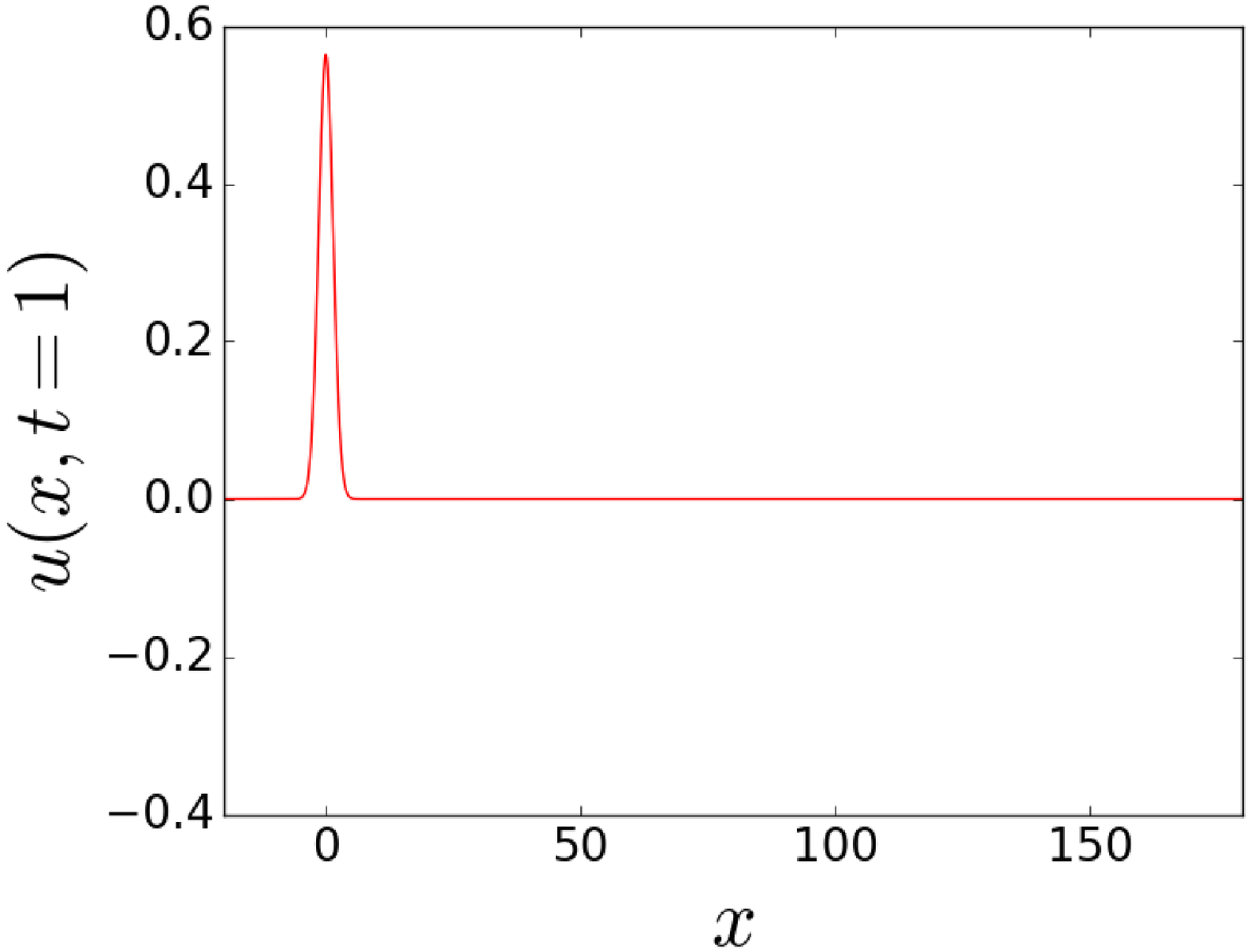}
			\caption{At~$t=1$}
			\label{6_000}
		\end{subfigure}%
		\begin{subfigure}{.34\textwidth}
			\centering
			\includegraphics[scale=0.28]{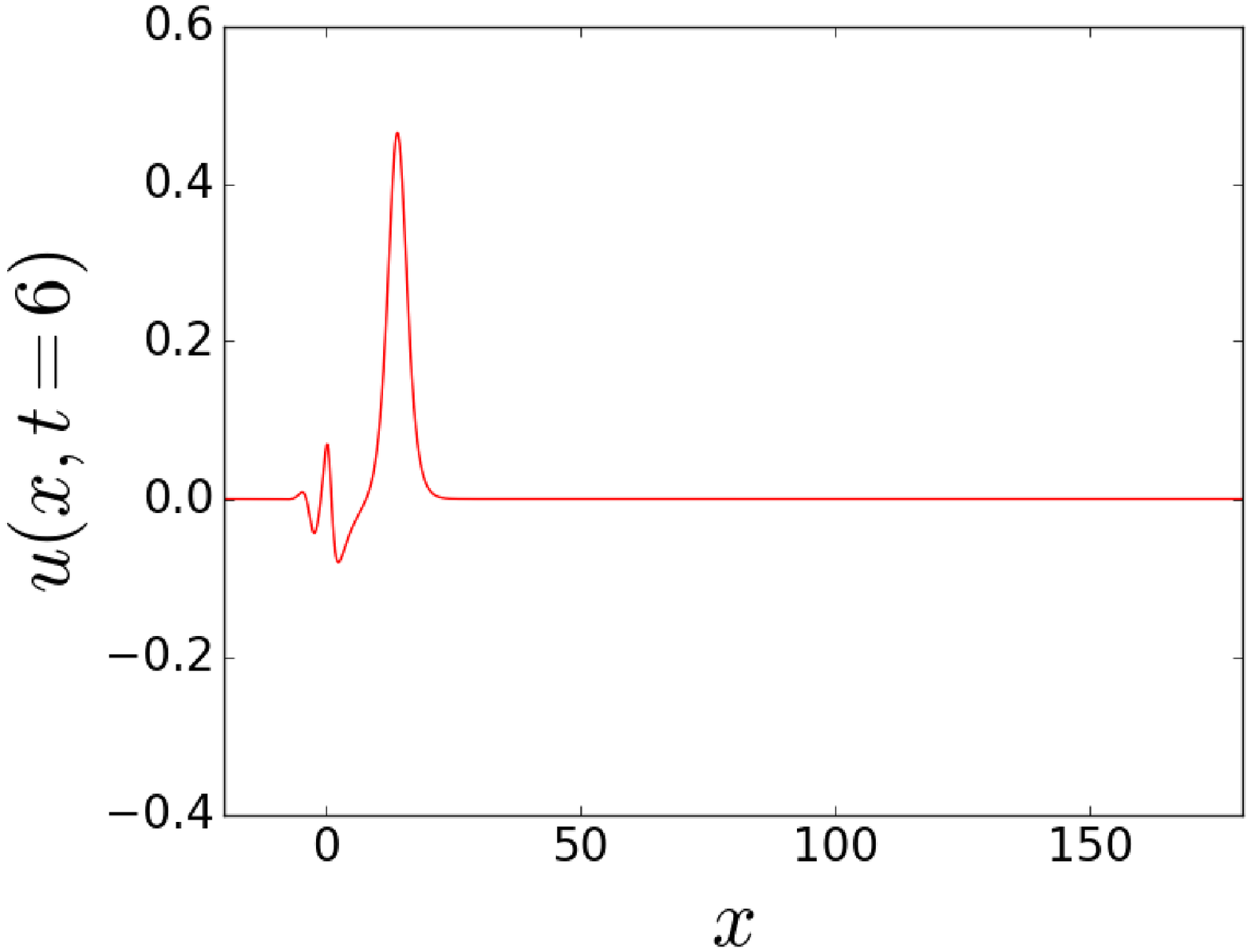}
			\caption{At~$t=6$}
			\label{6_040}
		\end{subfigure}%
		\begin{subfigure}{.34\textwidth}
			\centering
			\includegraphics[scale=0.28]{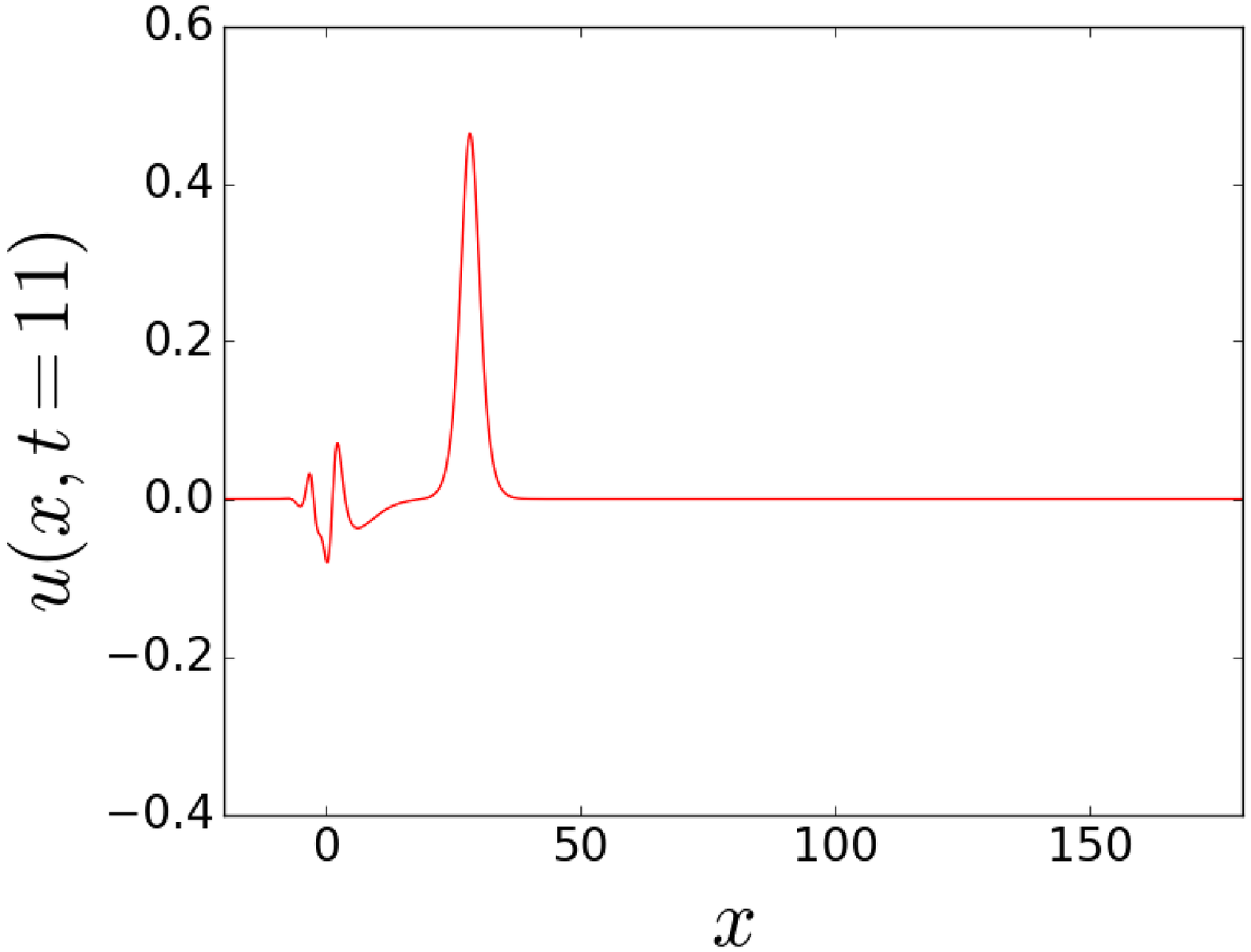}
			\caption{At~$t=11$}
			\label{6_080}
		\end{subfigure}
		
		\centering
		\hspace*{-0.1cm}
		\begin{subfigure}{.34\textwidth}
			\centering
			\includegraphics[scale=0.28]{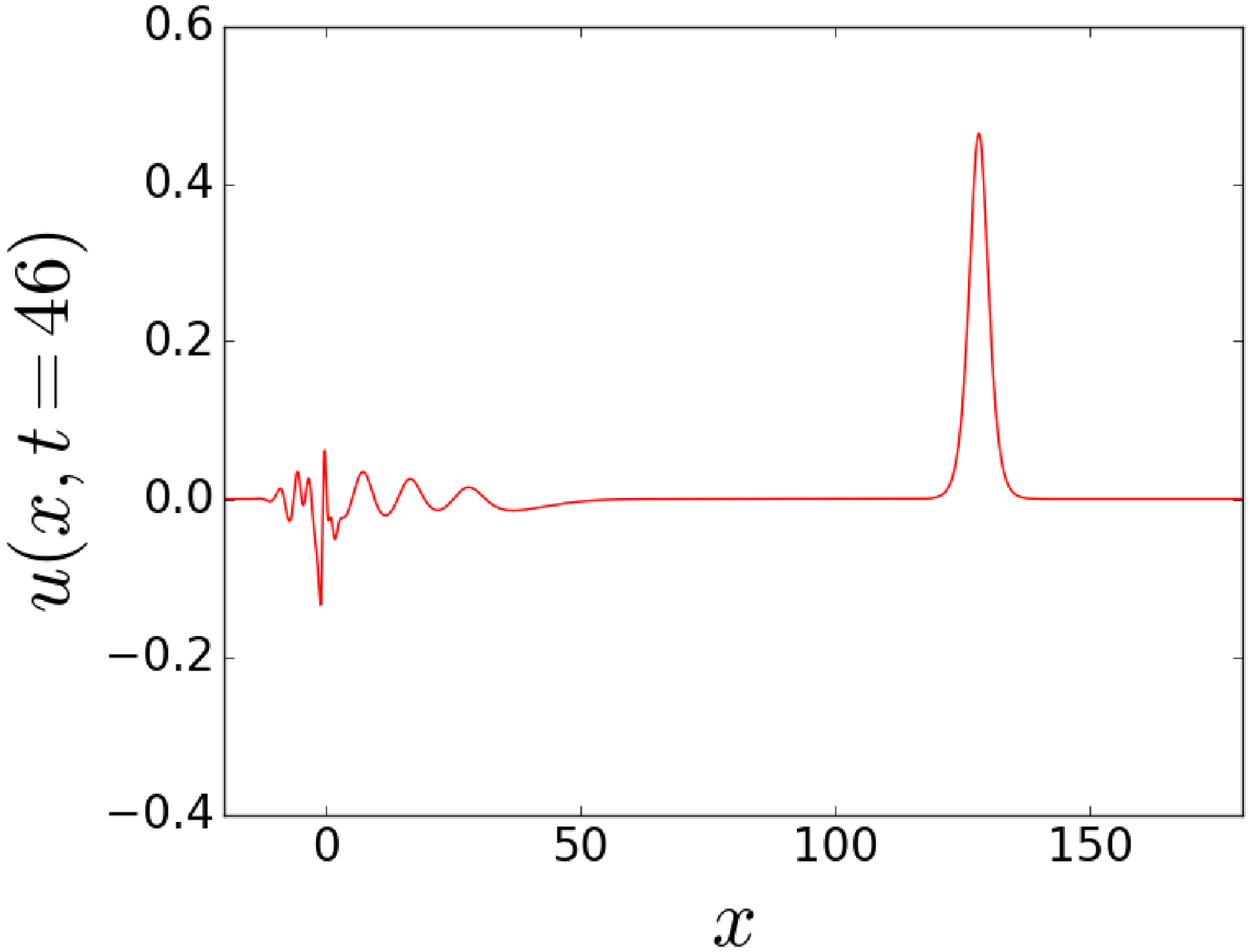}
			\caption{At~$t=46$}
			\label{6_120}
		\end{subfigure}%
		\begin{subfigure}{.34\textwidth}
			\centering
			\includegraphics[scale=0.28]{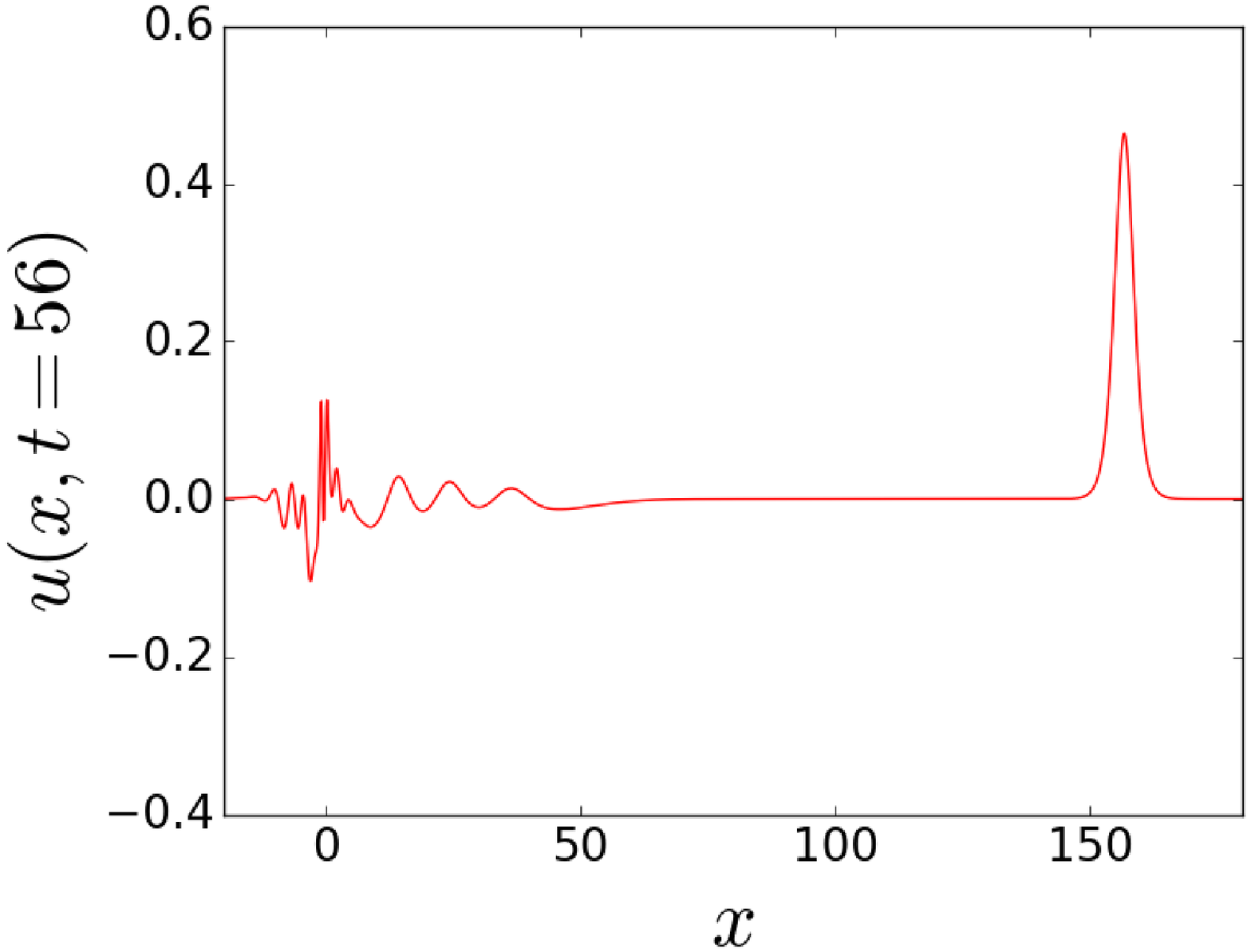}
			\caption{At~$t=56$}
			\label{6_160}
		\end{subfigure}%
		\begin{subfigure}{.34\textwidth}
			\centering
			\includegraphics[scale=0.28]{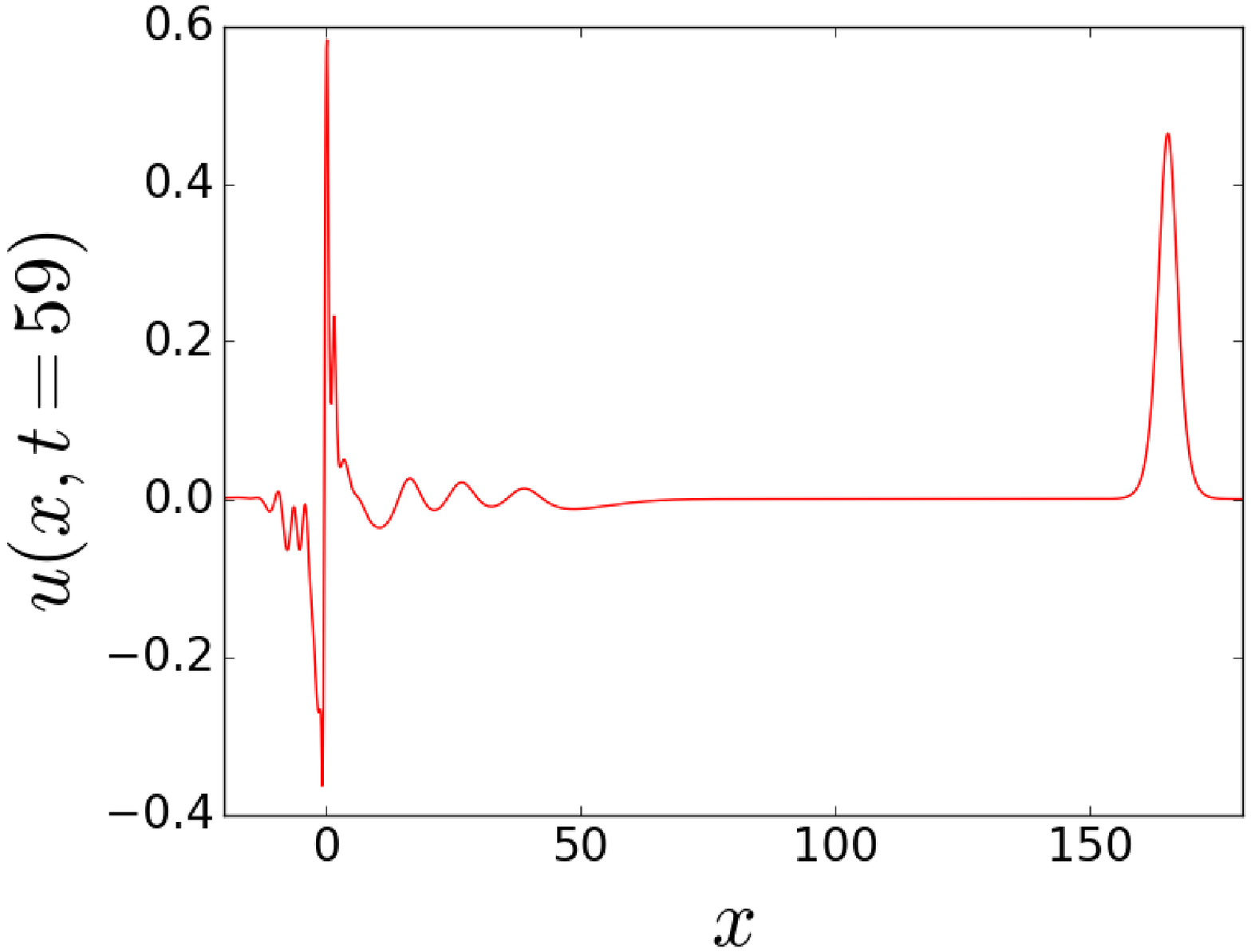}
			\caption{At~$t=59$}
			\label{6_200}
		\end{subfigure}
		\caption{Plots of $x$-dependence of $u(x,t)$ (at various values of $t$) of the numerical evolution of $q$ field
			with the initial condition taken from equation~(\ref{2.1}).}
		\label{plot6_000to6_200}
	\end{figure}

	However, this was not the case for the evolution of another `lump-like'
configuration, this time described by the following Gaussian function
	\begin{equation}
	q = -t \erf(0.5 x) \implies u = \frac{1}{\sqrt{\pi}} \exp\left(-x^2\right) \,. \label{2.1}
	\end{equation}
	Its time evolution is presented in figure~\ref{plot6_000to6_200}.
	This time we saw that the lump, which at the start of the simulation ($t=1$) was located at~$x=0$
(see figure~\ref{6_000}), started to move to the right, also leaving some `radiation' behind.
 However, our plots show that the `radiation' started to develop larger amplitudes as the time progressed. 
This can be seen by comparing figure~\ref{6_160} with figure~\ref{6_200} which show that the amplitudes of the
 `radiation' had grown significantly during a short period of time. In fact, at approximately $t=67.2$
 the whole system blew up. The same happened when we repeated the simulation with a negative amplitude 
({\it i.e.}, using $u = -\pi^{-1/2} \exp\left(-x^2\right)$ as initial conditions). In this case the simulation
 blew up at approximately $t =11.9$ ({\it i.e.} again, smaller than for a positive initial amplitude of the lump).

	Let us add that we believe that the blow-ups described in this section are not numerical artefacts,
 since we checked this by changing the parameters of the grid-spacing and time-steps, and they have always 
happened at roughly the same values of $t$.
		
	So, to sum up, we feel that the blow-ups described in this section imply 
that the soliton resolution conjecture does not hold for the mRLW equation, whereas it does hold for many
 integrable systems such as the KdV and the non-linear Schr\"{o}dinger equation. 
Numerically, this is the only observation we have found that distinguishes the non-integrable mRLW equation 
from many other integrable systems and this may be related to the fact that this model does not possess
a conserved quantity which controls and limits the growth of the amplitude (such as the energy in
 many systems).
	
	\section{Conclusions and further comments}
	  
	In this paper we have investigated the (numerical) time evolution 
of two- and three-soliton configurations of the mRLW equation. When we numerically evolved the 
initial conditions described by the analytic two-soliton solution and compared the
 results with the corresponding analytical simulation, we have found that they were essentially indistinguishable. This provided a good test of our numerical procedure and reassured us that the results of our simulations could be trusted. 
	Furthermore, our results agreed with the results presented in the original paper by Gibbon \textit{et al.}~\cite{Gibbon}, where they overlapped. In addition to the investigation of the numerical two-soliton solutions, we have also studied 
the numerical time evolution of systems constructed by the superposition of two one-soliton solutions, 
and have found that these configurations approximated the analytical two-soliton solutions very closely. 
	
	This has led us to consider three-soliton configurations constructed by taking
 a superposition of three one-soliton solutions. The numerical time evolution of these three-soliton 
configurations behaved very much like those seen in integrable models; the field evolved well,
 there did not seem to be any breaking or changes to the individual solitons whenever they were far from each 
other, and after the scattering the solitons emerged with their original shapes and velocities. Furthermore, 
the phase-shift experienced by each soliton was additive, and the known conservation laws were also obeyed 
for such configurations. This suggests to us that analytical three-soliton solutions may exist (though
 their analytic forms cannot be found by Hirota's method). 
	  
	We also looked at the time evolutions of various lumps - {\it i.e.}, fields that crudely resemble
 a single soliton field but are not exact solutions of the mRLW equation. We have found that the 
system blows up for some lumps, which is most likely a consequence of instabilities of the model which lead
to the development of very steep gradients causing our numerical procedures to break down. 
We have checked that these blow-ups were genuine properties of the evolution of these field configurations
 rather than numerical artefacts.
 Our results imply that the soliton resolution conjecture does not hold for the mRLW equation.
 Numerically, this is the only property of this non-integrable model with a behaviour that differs from 
many other integrable models. It would therefore be interesting to study other systems which are not Hirota 
integrable but do possess  one- and two-soliton Hirota solutions and check if such systems show 
similar properties to the mRLW equation ({\it i.e.},  whether they possess numerical three-soliton 
solutions in which the solitons behave exactly as one would expect from integrable solitons, but 
the soliton resolution conjecture does not hold). This could shine some new light on the connection 
between (Hirota) integrability and the soliton resolution conjecture. 
		
	So overall, our results show that the mRLW model has many interesting properties
 and in many ways behaves like an integrable model. This has led us to consider
whether we can think of this model as a finite perturbation of an integrable model
 and an obvious suggestion here is to think of this model as a perturbation of the KdV 
equation.\footnote{We thank L. A. Ferreira for this suggestion.} So together with L. A. Ferreira 
we are now trying to carry out such a procedure and we hope  that it will help us to understand 
the integrability/quasi-integrability properties of this model better.
 We hope to be able to say more on this topic soon.
	  
	{\bf Acknowledgements:} {We would like to thank L. A. Ferreira for
 many constructive and helpful comments and for working with us on some problems discussed 
in this paper. WJZ would like to thank the Royal Society for its grant to collaborate with L. A. Ferreira. Both authors thank the FAPESP/Durham University for their grant to facilitate their visits to the USP at S\~ao Carlos, and the Department of Physics in S\~ao Carlos for its hospitality.}

	\appendix
		
	\section{Numerical methods} \label{Numerical_method}
	
	Our numerical procedure for solving the mRLW equation combines the explicit and implicit methods of solving the equations of motion of the model. This is due to the fact that the term in the equation which contains the highest time derivatives also possesses spatial derivatives. A similar problem was encountered by J. C. Eilbeck and G. R. McGuire when they numerically investigated the regularized long-wave equation~\cite{Eilbeck}. Their paper provides a detailed discussion of how to use such methods. We followed their ideas, modifying them appropriately for our investigations, and we present a short discussion of our procedure in the following subsection. 
	
	\subsection{Numerical approximation of mRLW equation} \label{Numerical_approx}	
	
	First, to simplify the mRLW equation, we introduce a new field~$p$ as follows
	\begin{equation}
		p = q_t \label{A.1}
	\end{equation}
	so that the mRLW equation takes the form
	\begin{equation}
		p_{xxt} + 2 q_{xx} p_t + 4 p_x^2 - p_x - p_t = 0 \,. \label{A.2}
	\end{equation}
	We then take a finite set of points~$x_0, x_1, \ldots, x_N$ and~$t_0, t_1, \ldots, t_K$, 
and let~$h$ denote the grid-spacing and~$\tau$ the time-step. Furthermore, we denote the 
grid points as~$(i h , m \tau) \equiv (i,m)$, where~$i = 0,1,2,\ldots, N$ and~$m = 0,1,2,\ldots, K$, and we
 employ the following notation:~$p_i{}^m \equiv p(ih,m \tau)$ and~$q_i{}^m \equiv q(ih,m \tau)$. Finally, we choose~$v_i{}^m$ to denote our approximation to~$p_i{}^m$, and~$w_i{}^m$ to denote our approximation to~$q_i{}^m$. 
		
	Next, we introduce the following central finite difference operators by their actions on~$v_i{}^m$ 
as follows
	\begin{align}
		\delta_x^2 v_i{}^m & = (v^m_{i+1} - 2 v_i{}^m + v^m_{i-1})/h^2 \,, \\
		H_x v_i{}^m & = (v^m_{i+1}-v^m_{i-1}) / 2h \,, \\
		H_t v_i{}^m & = (v^{m+1}_i-v^{m-1}_i) / 2 \tau \,,
	\end{align}
	and similarly on~$w_i{}^m$. Applying these operators to equation~(\ref{A.2}) in a straightforward manner yields
	\begin{equation}
		\delta_x^2 H_t v_i{}^m + 2 \delta_x^2 w_i{}^m H_t v_i{}^m + 4 \left( H_x v_i{}^m \right)^2 - H_x v_i{}^m - H_t v_i{}^m = 0 \,,
	\end{equation}
	which can be rewritten as
	\begin{equation}
		\begin{aligned}
			\begin{split}
				- \frac{v_i^{m+1}}{2 \tau} & + \frac{v_{i+1}^{m+1}-2v^{m+1}_i + v_{i-1}^{m+1}}{2h^2 \tau} + \frac{w_{i+1}^m v_i^{m+1} - 2 w_i{}^m v_i^{m+1} +  w^m_{i-1}v_i^{m+1}}{h^2 \tau} \\& = - \frac{v_i^{m-1}}{2 \tau}  + \frac{v_{i+1}^m- v_{i-1}^m}{2h} + \frac{v_{i+1}^{m-1}-2v_i^{m-1}+v_{i-1}^{m-1}}{2h^2 \tau}  \\& \hspace{4.5 mm} + \frac{w_{i+1}^m v_i{}^{m-1} - 2w_i{}^m v_i^{m-1} + w_{i-1}^m v_i^{m-1}}{h^2 \tau}  \\& \hspace{4.5 mm} - \frac{(v_{i+1}^m)^2 - 2 v_{i+1}^m v_{i-1}^m + (v_{i-1}^m)^2}{h^2} \,. \label{A.2.1}
			\end{split}
		\end{aligned}
	\end{equation}
	Let us now introduce the following matrices
	\begin{equation}
		A \equiv \begin{pmatrix} 1 & 0 & 0 & \cdots &  & 0 \\
		\frac{1}{2h^2 \tau} & a_1{}^m & \frac{1}{2h^2 \tau} & 0 & & \vdots \\
		0 & \frac{1}{2h^2 \tau} & a_2{}^m & \frac{1}{2h^2 \tau} & \\
		\vdots &  & \ddots & \ddots & \ddots & 0 \\
		&  &  & \frac{1}{2h^2 \tau} & a_{N-1}^m & \frac{1}{2h^2 \tau} \\
		0 & \cdots & & 0 & 0 & 1  
		\end{pmatrix} \,,
	\end{equation}
	\begin{equation}
		B \equiv \begin{pmatrix} v_0^{m+1} \\ v_1^{m+1} \\ v_2^{m+1} \\ \vdots \\ v_{N-1}^{m+1} \\ v_N^{m+1} \end{pmatrix} \,,\quad  C \equiv \begin{pmatrix} c_0{}^{m} \\ c_1{}^{m} \\ c_2{}^{m} \\ \vdots \\ c_{N-1}^{m} \\ c_N^{m} \end{pmatrix} \,,
	\end{equation}
	where
	\begin{equation}
		a_i{}^m \equiv \frac{-1 + w_{i+1}^m - 2 w_i{}^m +  w_{i-1}^m}{h^2 \tau} - \frac{1}{2\tau} \,,
	\end{equation}
	\begin{equation}
		c_0{}^{m} \equiv v_0^{m+1} \,,\quad c_N^{m} \equiv v_N^{m+1}\,,
	\end{equation}
	\begin{equation}
		\begin{aligned}
			\begin{split}
				c_i{}^m \equiv & - \frac{v_i^{m-1}}{2 \tau} + \frac{v_{i+1}^m- 	v_{i-1}^m}{2h} + \frac{v_{i+1}^{m-1}-2v_i^{m-1}+v_{i-1}^{m-1}}{2h^2 \tau} \\& + \frac{w_{i+1}^m v_i{}^{m-1} - 2w_i{}^m v_i^{m-1} + w_{i-1}^m v_i^{m-1}}{h^2 \tau} \\& - \frac{(v_{i+1}^m)^2 - 2 v_{i+1}^m v_{i-1}^m + (v_{i-1}^m)^2}{h^2} \,,\quad i = 1, 2 , \hdots, N-1 \,,
			\end{split}
		\end{aligned}
	\end{equation}
	so that equation~(\ref{A.2.1}) can be rewritten as the matrix equation
	\begin{equation}
		AB = C \,.
	\end{equation}
	Hence, we need to solve this equation for the vector~$B$. This can done by using the well-known~$LU$~decomposition method~\cite{Schay}.
		
	Once we have solved for~$B$, we have the values of $v_i^m$ at the next time level. One can then find the values of all~$w_i^m$ by solving equation~(\ref{A.1}) using the central difference operator, that is,
	\begin{equation}
		v_i{}^m = \frac{w^{m+1}_i - w_i^{m-1}}{2 \tau} \implies w_i^{m+1} = 2 \tau v_i{}^m + w_i^{m-1} \,. \label{A.3}
	\end{equation}
	We then repeat this procedure for all time levels in order to calculate the time evolution of any initial configuration.
		
	It is not too difficult to verify that this scheme is both second-order accurate in~$\tau$ and in~$h$. Furthermore, we have extensively tested this scheme against the analytical one- and two-soliton solutions. These tests have shown that the numerical simulations approximate the analytical solutions extremely well without any (visible) loss of radiation. This indicates that, for the values of $h$ and $\tau$ that we have used, the scheme is stable and we can trust its results.
		
	Finally, let us add that had we substituted the finite difference operators directly into the
 mRLW equation without first making the substitution expressed by equation~(\ref{A.1}),
 then the~$4q_{xt}^2$ term would have yielded a term~$(-w_{i+1}^{m+1}w_{i-1}^{m+1})/(2h^2 \tau^2)$; and
 since this term contains two unknowns, ({\it i.e.},~$w_{i+1}^{m+1}$ and~$w_{i-1}^{m+1}$), we would 
not have been able to solve for it. Moreover, for very similar reasons, we cannot use the well-known Crank-Nicolson method to numerically solve the mRLW equation.
		
		
		
	\subsection{Summary of parameters used to produce figures} \label{summary_of_variables}
		
	Table~\ref{variables1} shows all the parameters used to produce the figures shown in this paper.
	\begin{table}[H]
		\center
		\caption{This table summarizes the variables used on the simulations discussed in this paper and presented as plots in various figures.}
		\begin{tabular}{ l?{0.5mm} c | c | c | c | c}
			\toprule[1.5pt]
			& Figure~\ref{plot0_2to0_7} & Figures \ref{plot1t200to1t675} and \ref{1amplitude_vs_time_and_1location_vs_time} & Figures \ref{plot3t200to1t775} and \ref{3amplitude_vs_time_and_1location_vs_time} & Figure \ref{plot8_000to6_250} & Figure~\ref{plot6_000to6_200}  \\
			\midrule
			$x_0$ & $-50$ & $-220$ & $-125$ & $-600$ & $-150$ \\
			$x_N$ & $250$ & $250$ & $175$ & $1200$ & $350$ \\
			$h$ & $0.1$ & $0.1$ & $0.1$ & $0.1$ & $0.1$ \\
			$t_0$ & $0$ & $0$ & $0$ & $1$ & $1$ \\
			$t_K$ & $40$ & $170$ & $42$ & $241$ & $81$ \\
			$\tau$ & $0.001$ & $0.001$ & $0.001$ & $0.01$ & $0.001$ \\
			$\omega_1$ & $5.00$ & $0.80$ & $0.80$ & N/A & N/A \\
			$\delta_1$ & $0.00$ & $66.51$ & $16.63$ & N/A & N/A \\
			$\omega_2$ & $3.00$ & $1.33$ & $3.07$ & N/A & N/A \\
			$\delta_2$ & $0.00$ & $110.90$ & $63.77$ & N/A & N/A \\
			$\omega_3$ & N/A & $1.84$ & $4.28$ & N/A & N/A \\ 
			$\delta_3$ & N/A & $152.88$ & $89.00$ & N/A & N/A
			\label{variables1}
		\end{tabular}
	\end{table}

\end{document}